\documentclass[aps,prl,reprint,superscriptaddress,nofootinbib,nobibnotes,longbibliography]{revtex4-2}

\usepackage{amsmath,amssymb,bm,mathtools}
\usepackage{braket}
\usepackage{graphicx}
\usepackage{xcolor}
\usepackage{microtype}
\usepackage{tikz}
\usetikzlibrary{arrows.meta,calc,fit,positioning,shapes.geometric}
\usepackage[qm]{qcircuit}
\usepackage[hidelinks]{hyperref}
\hypersetup{colorlinks=true,citecolor=blue,
linkcolor=blue,urlcolor=blue,pdfstartview=FitH,
bookmarksopen=true}

\definecolor{dreamblue}{RGB}{38,92,150}
\definecolor{dreamred}{RGB}{176,55,51}
\definecolor{dreamgreen}{RGB}{50,125,92}
\definecolor{dreamgold}{RGB}{190,139,33}
\definecolor{dreamgray}{RGB}{88,94,103}

\newcommand{\F}{\mathcal F}
\newcommand{\STAB}{\mathrm{STAB}}
\newcommand{\SEP}{\mathrm{SEP}}
\newcommand{\LSCC}{\mathrm{LSCC}}
\newcommand{\id}{\mathbb I}
\newcommand{\Tr}{\operatorname{Tr}}

\newcommand{\cnot}{\mathrm{CNOT}}
\newcommand{\ketbra}[1]{\ket{#1}\!\bra{#1}}

\newtheorem{theorem}{Theorem}
\newtheorem{proposition}[theorem]{Proposition}
\newtheorem{lemma}[theorem]{Lemma}
\newtheorem{corollary}[theorem]{Corollary}

\begin{document}

\title{Distributed Resource Theory of Entanglement and Magic}

\author{Xiao Yuan}
\email{xiaoyuan@pku.edu.cn}

\author{Wenhao Zhang}
\email{1701210129@pku.edu.cn}
\author{Qiming Ding}
\email{dqiming94@pku.edu.cn}
\affiliation{Center on Frontiers of Computing Studies, Peking University, Beijing 100871, China}
\affiliation{School of Computer Science, Peking University, Beijing 100871, China}

\author{You Zhou}
\email{you\_zhou@fudan.edu.cn}
\affiliation{Key Laboratory for Information Science of Electromagnetic Waves (Ministry of Education), Fudan University, Shanghai 200433, China}


\begin{abstract}
In distributed fault-tolerant quantum computing, entanglement and magic are essential resources for quantum communication and universal fault-tolerant computation, respectively. Although they are usually treated as distinct resource currencies, whether they admit a unified resource-theoretic description remains an open question. Here, we introduce the distributed resource theory of entanglement and magic (DREAM). In this framework, the free states are convex mixtures of product local stabilizer states, and the free operations are local stabilizer circuits assisted by classical communication (LSCC). We show that DREAM contains nontrivial resource states that are neither entanglement nor magic, so it is strictly richer than treating the two resources independently. 
Surprisingly, such resources can enable the teleportation of magic states between distant parties without consuming entanglement, revealing a counterintuitive form of resource teleportation mediated entirely by separable states. We further generalize this result to quantum networks and investigate general quantum-state teleportation under LSCC. We show that the shared resource under DREAM is closely related to the teleportation capability, quantified by the magic of the teleported state. Our work establish a systematic framework for studying distributed quantum resources and uncover intrinsic relations among distinct resources within a unified resource theory.
\end{abstract}

\maketitle

Quantum computing is advancing rapidly, yet scaling up monolithic quantum processors faces severe physical bottlenecks, such as fabrication defects, wiring congestion, and cross-talk during error correction~\cite{Awschalom2021}. A compelling strategy to overcome these constraints is distributed quantum computing (DQC), which networks multiple moderate-sized quantum processing modules via classical and quantum communication channels~\cite{Briegel1998,Monroe2014,Nickerson2014,Kimble2008,Wehner2018,Cuomo2020,VanMeter2014,nickerson2013topological}. By coordinating local processors, DQC executes large-scale quantum circuits without compromising performance or qubit connectivity~\cite{Gottesman1999,Cirac1999,Jiang2007}. Recent experimental breakthroughs have demonstrated the practical feasibility of this architecture; for instance, deterministic distributed quantum algorithms and quantum gate teleportation have been realized across photonically linked modules~\cite{Main2025}. By shifting the scaling challenge from fabricating ultra-dense single chips to manufacturing modular units and coherent quantum interfaces, DQC provides a scalable pathway toward fault-tolerant quantum computation and naturally fits the architecture of quantum networks.

Two fundamental resources are indispensable in DQC. First, non-local entanglement~\cite{Peres1996,Horodecki1997,vedral1997quantifying,vidal2000entanglement,plenio2007introduction,Horodecki2009,Knill1998,Ollivier2001,Cubitt2003}, in the ideal form of shared Bell pairs,  is the essential currency for inter-module quantum communication, enabling quantum state teleportation~\cite{Bennett1993}, remote gate execution~\cite{eisert2000optimal}, and lattice-surgery primitives~\cite{Horsman2012,Fowler2012,litinski2019game}. Second, magic states~\cite{Gottesman1998,bravyi2005universal,Veitch2014,HowardCampbell2017,leone2022stabilizer}, typically prepared and distilled locally in dedicated nodes such as magic-state factories~\cite{bravyi2005universal,litinski2019game}, provide the non-stabilizer resources required to implement non-Clifford operations for universal quantum computation~\cite{Gottesman1998}. These two ingredients are currently governed by mature yet separate resource theories~\cite{Chitambar2019}: entanglement theory treats separable states as free under local operations and classical communication (LOCC)~\cite{vidal2000entanglement}, whereas magic-state theory treats the stabilizer polytope as free under stabilizer operations~\cite{BravyiSmithSmolin2016,Beverland2020}. Because distributed fault-tolerant architectures simultaneously impose spatial partitioning and stabilizer restrictions~\cite{Beverland2022,Bravyi2019,HouCaoYang2026}, evaluating entanglement and magic in isolation fails to capture the intrinsic resource structure. This raises a deeper question: are there resources beyond entanglement and magic that are more fundamental to distributed quantum computing?

Here we address the questions by introducing a distributed resource theory of entanglement and magic (DREAM), a unified framework for DQC. First, we identify quantum states that possess neither entanglement nor magic, yet remain resourceful within DREAM, thereby revealing a distinct type of resource beyond conventional entanglement and magic. We show that these states are not merely of theoretical interest: they can be used to teleport magic states between two distant parties. This leads to a counterintuitive form of resource teleportation that requires \emph{no} entanglement. We then extend the result to general quantum networks, uncovering an intimate relation among entanglement, magic, and the newly identified resource. We further study the more general task of quantum-state teleportation between two parties and derive an analytical relation between the required resource under DREAM and the magic of the teleported state. Together, these results establish a systematic framework for understanding the intrinsic roles and interplay of quantum resources beyond conventional entanglement and magic in DQC.

\vspace{0.2cm}

\emph{Framework.---}We consider a quantum network described by a finite graph $G=(V,E)$, where the vertices $V$ represent local quantum systems and the edges $E$ represent communication links between them.

We first recall the resource-theoretic descriptions of entanglement and magic. For entanglement, the set of free states is the set of fully separable states,
$
    \SEP_V
    =
    \operatorname{conv}\!\left\{
    \bigotimes_{v\in V}\sigma_v:
    \sigma_v\in\mathcal{D}_v
    \right\}$,
where $\mathcal{D}_v$ denotes the set of density operators on node $v$. The free operations are local operations and classical communication (LOCC). Entanglement therefore captures quantum correlations that cannot be generated between spatially separated nodes using LOCC alone. For magic, the free-state set $\STAB_V$ is the stabilizer polytope of the composite system, and the free operations are stabilizer operations~\cite{HowardCampbell2017}. Magic quantifies the nonstabilizer resource required to implement universal quantum computation beyond Clifford operations.

We now turn to distributed fault-tolerant quantum computing, where both entanglement and magic constitute valuable resources, and introduce the distributed resource theory of entanglement and magic (DREAM). We define the set of free states in DREAM as
\begin{equation}    \label{eq:free}
    \F_V
    =
\operatorname{conv}\!\bigg\{
    \bigotimes_{v\in V}\sigma_v:
    \sigma_v\in\STAB_v
    \bigg\},
\end{equation}
where $\STAB_v$ denotes the stabilizer polytope on node $v$. Thus, a free state in DREAM is a convex mixture of product states whose local components are stabilizer states. Correspondingly, the free operations are local stabilizer operations and classical communication (LSCC), namely, finite-round protocols generated by node-local stabilizer-state preparation, Clifford isometries, Pauli measurements, discarding of subsystems, shared randomness, classical feedforward, and classical communication between nodes.

By construction, $\F_V\subseteq\SEP_V$ and $\F_V\subseteq\STAB_V$: every free state in DREAM is both separable across the network and globally nonmagical. Importantly, however, the converse need not hold. As we show below, the structure of $\F_V$ is richer than what follows from simply combining the conventional resource theories of entanglement and magic. Although the above definition may arise naturally from the general framework of quantum resource theories~\cite{Chitambar2019,Ganardi2026}, it gives rise to qualitatively new resource phenomena specific to distributed quantum computation.

\vspace{0.2cm}

\emph{Beyond entanglement and magic.---}We first show that DREAM contains resources that are captured by neither entanglement nor magic. Specifically, the  free-state set of DREAM is a strict subset of the states that are simultaneously separable and globally stabilizer:
\begin{equation}
    \F_V \subsetneq \SEP_V \cap \STAB_V .
    \label{eq:strict_inclusion}
\end{equation}
Thus, a state can be free with respect to both conventional resource theories individually, yet remain resourceful under DREAM.

To demonstrate this separation explicitly, we consider a bipartite system. Define
$
    M=(X+Y)/{\sqrt{2}}$ with $X$ and $Y$ the corresponding Pauli operators,
    the states
    $\tau_\pm=(\id\pm M)/{2}$,
and the one-parameter family
$
    \rho_\eta
    =
    \frac{1}{4}
    \left(\id_4+\eta M\otimes M\right),\, |\eta|\leq 1$.
At $\eta=1$, the state admits two equivalent decompositions
\begin{equation}    \label{eq:twodecomp}
\begin{aligned}
    \rho_1
    &=
    \frac{1}{2}
    \left(
    \tau_+\otimes\tau_+
    +
    \tau_-\otimes\tau_-
    \right),\\
    &=
    \frac{1}{2}
    \left(
    \ketbra{\Psi^+}
    +
    \ketbra{\Phi_i}
    \right),
\end{aligned}
    \end{equation}
where
$\ket{\Psi^+}=(\ket{01}+\ket{10})/\sqrt{2}$ and
$\ket{\Phi_i}=(\ket{00}+i\ket{11})/\sqrt{2}$ are two stabilizer states, which are equivalent to the standard Bell pair under local Clifford gates.
The first decomposition shows that $\rho_1$ is separable, whereas the second shows that it lies in the global stabilizer polytope. Hence, $\rho_1$ possesses neither entanglement nor magic. Note that the same argument holds for the entire family $\rho_\eta$.

Despite being free under both conventional resource theories, $\rho_\eta$ is not always free under DREAM. In fact, we prove the following exact characterization:
\begin{theorem}
\label{thm:geometry} 
    $\rho_\eta$ is free under DREAM iff $|\eta|\le \frac{1}{2}$.
\end{theorem}
Consequently, $\rho_\eta$ belongs to $
    \left(\SEP_{AB}\cap\STAB_{AB}\right)
    \setminus\F_{AB}$ for $|\eta|>1/2$,
providing an explicit separation between DREAM and the individual resource theories of entanglement and magic.
The separation can be quantified by the standard robustness~\cite{vidal1999robustness,takagi2019general},
$R_{\rm s}^{\mathcal F_V}(\rho_\eta)
    =
    \max\left\{0,|\eta|-\frac{1}{2}\right\}$,  with $R_{\rm s}^{\F}(\rho)=\min\left\{s\ge0:{\rho+s\sigma_-}{}=(1+s)\sigma_+,
 \ \sigma_\pm\in\F\right\}$.
Equivalently, $M\otimes M$ serves as a witness: its expectation value on every product stabilizer state is bounded in magnitude by $1/2$, whereas
$    \Tr\!\left(\rho_\eta M\otimes M\right)=\eta$. We refer to SM for details

Interestingly, this resource cannot be reduced to either conventional resource individually. Since
$\rho_\eta\in\SEP_{AB}\cap\STAB_{AB}$, 
any copies of $\rho_{\eta}$ cannot be converted under LSCC into either an entangled state or a magical state, even probabilistically on a nonzero heralded branch. The strict inclusion in Eq.~\eqref{eq:strict_inclusion} therefore identifies a genuine distributed resource sector that remains invisible to both entanglement and magic separately.
In SM, we further extend this construction to multipartite networks and show that such resources can be genuinely distributed while remaining simultaneously separable and globally stabilizer state. 
These results reveal a nontrivial resource structure intrinsic to DREAM that goes fundamentally beyond a simple combination of entanglement and magic.

\vspace{0.2cm}

\emph{Resource teleportation without entanglement.---}While the above results show that DREAM goes beyond entanglement and magic at the level of resource-theoretic structure, a more operational question is whether these resources have a concrete information-processing role.
Surprisingly, we find that the resource state defined in Eq.~\eqref{eq:twodecomp}, despite being separable, can be used to teleport a magic state between two distant parties.

\begin{theorem}
There exists an LSCC protocol: $ \tau_+^{a}\otimes\rho_1^{AB}
 \xrightarrow[\text{one cbit}]{\LSCC}\tau_+^{B}$.
\label{thm:perfect}
\end{theorem}

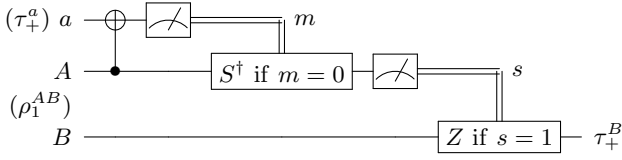
\begin{figure}[t]
\centering
\mbox{
\Qcircuit @C=0.85em @R=0.65em {
 \lstick{(\tau_+^a)\ a} & \targ & \meter    &\rstick{m}\cw\\
  \lstick{A} & \ctrl{-1} & \qw & \gate{S^{\dagger}\ {\rm if}\ m=0}\cwx[-1] & \meter   &\rstick{s}\cw \\
 \lstick{(\rho_1^{AB})}  \\
 \lstick{B} & \qw & \qw & \qw & \qw & \gate{Z\ {\rm if}\ s=1}\cwx[-2]  &\rstick{\tau_+^B}\qw
}}
\caption{Gate-level realization in time order.  The $Z_a$ measurement on system $a$ produces $m\in\{0,1\}$; Alice's final $X_A$ measurement on system $A$ produces the hidden sign $s\in\{0,1\}$.  Only $s$ is sent to Bob.}
\label{fig:Scircuit}
\end{figure}

\noindent
An explicit realization of the protocol is shown in Fig.~\ref{fig:Scircuit}. Alice applies $\cnot_{A\to a}$ to her two local systems, measures system $a$ in the $Z$ basis and obtain outcome $m$, and then measures $Y_A$ or $X_A$ depending on the outcome $m$. The induced operators $T/\sqrt2$ and $e^{i\pi/4}T^\dagger/\sqrt2$ realize an exact measurement of the hidden $\tau_\pm$ label in Eq.~\eqref{eq:twodecomp}. Bob then applies either $I$ or $Z$, conditioned on Alice's measurement outcome $s$, yielding the magic state $\tau_+$ on Bob's side.

Thus, the protocol consumes the shared separable state $\rho_1$ and transfers a magic state from Alice to Bob without using entanglement. This seems counterintuitive because entanglement is generally required for teleporting a general quantum state. In fact, there is no conflict with standard quantum-information principles: the state being transferred here is pre-fixed and known. If arbitrary local operations were available, Bob could trivially prepare the state locally rather than relying on teleportation.
Nevertheless, the distinction becomes operationally relevant in distributed fault-tolerant quantum computing, where different nodes may have different capabilities. For example, a server equipped with a magic-state factory may be able to prepare magic states locally, whereas remote users may be restricted to stabilizer operations. In this setting, one relevant task is not to transmit an unknown quantum state, but to transfer a magic state that cannot be created locally by the receiving user. DREAM naturally captures this scenario. The above protocol shows that the resource $\rho_1$ can enable such magic state transfer even in the complete absence of entanglement, providing an operational manifestation of the resource structure beyond entanglement and magic.

Furthermore, we define the maximal free weight
$$w_{\F_{AB}}(\rho)=\max\{\lambda:\rho=\lambda\sigma+(1-\lambda)\chi,
 \ \sigma\in\F_{AB},\ \chi\in \mathcal D_{AB} \}$$
as the maximum weight of a free component contained in $\rho$, which is similar to the definition of weight in the resource theory~\cite{Elitzur1992,Lewenstein1998,Bu2018,Yao2020,Ducuara2020,Uola2020,ding2022no}. We prove the following result.

\begin{corollary}
Perfect magic-resource teleportation,
$ \tau_+^{a}\otimes\omega^{AB}
 \xrightarrow[\text{one cbit}]{\LSCC}\tau_+^{B}$,
is possible only if $w_{\F_{AB}}(\omega^{AB}) = 0$.
\end{corollary}
This result further establishes the operational significance of the resource captured by DREAM. Intuitively, any nonzero free component of $\omega^{AB}$ remains free in every refined branch on Bob's side. Deterministic generation of the pure state $\tau_+$, together with its extremality, would therefore require the free component itself to produce $\tau_+$, which is impossible. Formally, this follows from the free-component no-purification principle~\cite{FangLiu2019,FangLiu2022,ding2022no}, once we establish that the induced map from the free input component to Bob is free-preserving.



\vspace{0.2cm}

\emph{Resource teleportation in fault-tolerant networks.---}Now we extend the result to the scenario of a general quantum network. We first consider the case where entanglement is also allowed.  Specifically, suppose Alice holds $q$ independent magic states and shares $c_\Phi$ Bell pairs and $h_{\rm b}$ resource $\rho_{\rm 1}$ with Bob, we have the following result. 

\begin{corollary}The maximum number of magic states that can be teleported from Alice to Bob is $\min\{q,c_\Phi+h_{\rm b}\}$
\label{cor:service}
\end{corollary}

\noindent The achievability is directly proved by combining the teleportation and the LSCC protocol in Theorem~\ref{thm:perfect}. We further prove the converse bound in SM.


The results generalize to the law of resource flow in a quantum network.  Let node $v$ initially hold $s_v$ perfect magic states and demand $d_v$ outputs.  Each undirected edge $e$ contains $c_e$ Bell pairs and $h_e$ independent $\rho_{\rm 1}$ resource, each consumable once in either direction.  For $R\subseteq V$, write $s(R)=\sum_{v\in R}s_v$ and $d(R)=\sum_{v\in R}d_v$, and let $\delta R$ denote the edge boundary of $R$. Then we can prove:

\begin{corollary}\label{cor:cut}
All demands are feasible under finite-round LSCC iff
$
 {d(R)\le s(R)+\sum_{e\in\delta R}(c_e+h_e),\
 \forall R\subseteq V.}$
\end{corollary}

\noindent This resource teleportation problem in a quantum network can be analyzed as a network flow problem by combining the magic teleportation protocol in Theorem~\ref{thm:perfect} and quantum teleportation via Bell pairs. The sum of Bell pairs and $\rho_1$ shared by one edge characterizes the capacity of this edge, i.e.~how many magic states could be transported through it. Necessity is a natural generalization of Corollary~4 when one takes the bi-separation $\{R,V\setminus R\}$ by regarding the global system as bipartite, composed of the two subsystems. Sufficiency can be proved using the max-flow min-cut theorem in network theory \cite{CLRS}. Details of the proof are provided in SM.

\vspace{0.2cm}

\emph{Teleporting a general state.---}Finally, we consider the more general task of teleporting an arbitrary quantum state rather than a fixed magic state. Suppose that a server, Alice, physically holds an $n$-qubit pure state $\psi^a$, and the goal is to transfer $\psi^a$ to a remote user, Bob, by consuming a shared resource state $\omega^{AB}$. If $\psi^a$ is completely unknown, standard quantum teleportation generally requires $n$ Bell pairs. Here, instead, we focus on the setting in which the classical description of $\psi^a$ is known. This naturally arises when $\psi^a$ is a prescribed resource state, such as a magic state or a CCZ state, or is generated by executing a delegated circuit requested by Bob. The relevant question is then: what distributed resource is required to transfer such a known quantum state under the DREAM framework?

First, we show that successful teleportation requires the net resource contained in the shared state $\omega^{AB}$ to be at least as large as the local resource, namely, the magic contained in $\psi^a$:
\begin{equation}
    R_{\rm s}^{\mathcal F}(\omega^{AB})
    \ge R_{\rm s}^{\rm STAB_A}(\psi^a).
\end{equation}
Thus, DREAM is a necessary resource for state teleportation whenever the target state possesses nonzero magic. However, although this bound holds generally, it is not necessarily tight. For example, for the protocol in Theorem~\ref{thm:perfect}, we have
$R_{\rm s}^{\mathcal F}(\rho_1^{AB})=1/2>
R_{\rm s}^{\rm STAB_A}(\tau_+^a)=(\sqrt{2}-1)/2$,
demonstrating a finite gap between the two quantities.

To obtain a more tight relationship, we need to notice the hidden resource contained in Alice's local state $\psi^a$. Specifically, when we apply an LSCC to $\psi^a\otimes\omega^{AB}$, the resource contained in $\psi^a$ could even help us realize a stronger teleportation protocol. In order to take into account such effects, we define a new set of free states
\begin{equation}
 \mathcal Q_{B|A}^{(\psi^a)}:=\operatorname{conv}\!\left\{
 \alpha_A\otimes\sigma_B:\
 \alpha_A\in\mathcal S_A,\ \sigma_B\in\STAB_B
 \right\},
 \label{eq:directed-free}
\end{equation}
where $\mathcal S_A = \STAB_A\cup \STAB_A(\psi^a)$ with $\STAB_A(\psi^a)$ being the set of states generated by applying stabilizer operations on $\psi^a$. Such free states include the resource contained in the to-be-teleported state $\psi^a$. Now, consider $\{\mathcal P_x\}_x$ as any probabilistic LSCC instrument, we have the output state 
$\Lambda_x^\psi(\omega^{AB}):=\Tr_{\rm rest}\mathcal P_x(\psi^a\otimes\omega^{AB})=p_x\zeta_x^B$ with probability $p_x$. The protocol succeeds when $\zeta_x^B = \psi^a$.

\begin{theorem}\label{thm:directed-transport}
Suppose a  probabilistic LSCC protocol exist to teleport $\psi^a$ using  $\omega^{AB}$ with probability $p$, then
\begin{equation}
  \begin{aligned}
    \label{eq:SstrongRg}
    p R_{\mathrm s}^{\STAB_A}(\psi^a)
    &\leq
    R_{\mathrm s}^{\mathcal Q_{B|A}^{(\psi^a)}}(\omega^{AB})
    \leq
    R_{\mathrm s}^{\F_{AB}}(\omega^{AB}).
  \end{aligned}
\end{equation}
\end{theorem}

\noindent We note that the result hold for deterministic protocols with $p=1$. The proof also applies to the general robustness measure. For the first half of the inequality, we could decompose $\omega$ into linear combination of a free state $q\in\mathcal Q_{B|A}^{(\psi^a)}$ and another state $\chi$, $\omega=(1+r)q-r\chi$, where $\chi$ is any free state for standard robustness (and arbitrary for general robustness). We then show that finite-round LSCC can only transform the free state into a stabilizer state on Bob's side, hence providing by feasibility an upper bound on the magic content of Bob's received state. The second half of each inequality is a direct consequence of $\F_{AB}\subseteq\mathcal Q_{AB}^{(\psi^a)}$. The details of the proof are provided in SM.

The inequality in Eq.~\eqref{eq:SstrongRg} establishes a much tighter relation between the magic of the teleported state $\psi^a$ and the DREAM resource consumed from $\omega^{AB}$. In particular, for the protocol in Theorem~\ref{thm:perfect}, we find
$R_{\mathrm s}^{\mathcal Q_{B|A}^{(\psi^a)}}(\omega^{AB})
=R_{\rm s}^{\rm STAB_A}(\tau_+^a)
=(\sqrt{2}-1)/2$,
showing that the bound is saturated and hence that the protocol is optimal.

\emph{Discussion.---}We have introduced DREAM, a resource-theoretic framework for jointly describing entanglement and magic in distributed fault-tolerant quantum computing. Its free states are convex mixtures of product local stabilizer states, with LSCC as the corresponding free operations. A central finding is that the DREAM free-state set is strictly smaller than the intersection of the separable and global stabilizer sets. Consequently, there exist distributed resources that possess neither entanglement nor magic individually, revealing a resource structure that cannot be captured by simply combining the two conventional theories.

Beyond this structural distinction, we have established an operational role for such resources. In particular, a separable and globally stabilizer state can mediate perfect transfer of a magic state between distant nodes without consuming entanglement. The maximal free weight further constrains when such perfect resource teleportation is possible. When entanglement and DREAM resources coexist, we obtain resource-flow bounds for magic-state distribution and extend them to general quantum networks through cut constraints. Finally, for teleportation of a general known quantum state, we derive quantitative bounds relating the resource contained in the shared state to the magic of the teleported state. By incorporating the resource already available in the input state, the directed resource construction further sharpens this relation and reveals how local magic and distributed resources cooperate in state transfer.

We envision DREAM as a starting point for a broader theory of distributed quantum resources, with several directions for future work toward uncovering new principles arising from the interplay among distinct quantum resources. A first question is to develop a more complete characterization of the additional distributed resource identified by DREAM, including its monotones, conversion rates, formation costs, and distillation properties. It is also important to extend the present exact results to approximate and noisy transformations, where finite-error and finite-copy tradeoffs should be directly relevant to fault-tolerant architectures. More generally, characterizing resource capacities and optimal routing on networks with heterogeneous nodes and links may lead to a resource theory of distributed fault-tolerant computation at the network level. Finally, connecting DREAM resource measures to concrete implementation costs---such as magic-state factories, logical qubits, communication links, and code parameters---could clarify how the fundamental resource relations uncovered here translate into practical architectural advantages. These directions may ultimately lead to general laws governing the conversion, transport, and distribution of quantum computational resources across fault-tolerant quantum networks.

\vspace{0.2cm}

\emph{Acknowledgment.---}We thank Bartosz Regula, Ryuji Takagi, and Mile Gu for insightful discussions. This work is supported by the Beijing Natural Science Foundation Z250004 and No.~1254053, 
the National Natural Science Foundation of China Grant (No.~12361161602), 
NSAF (Grant No.~U2330201), Quantum Science and Technology-National Science and Technology Major Project (2023ZD0300200), and 
Beijing Science and Technology Planning Project (Grant No.~Z25110100810000).
Y.Z. further acknowledges funding from NSFC Grant No.~12575012, Quantum Science and Technology-National Science and Technology Major Project Grant Nos.~2024ZD0301900 and 2021ZD0302000, the Shanghai QiYuan Innovation Foundation, the Shanghai Municipal Commission of Science and Technology with Grant No.~25511103200, the Shanghai Science and Technology Innovation Action Plan Grant No.~24LZ1400200, the Shanghai Pilot Program for Basic Research - Fudan University 21TQ1400100 (25TQ003), the CCF-Quantum CTek Superconducting Quantum Computing CCF-QC2025006.

The authors used OpenAI Codex as an AI-assisted research tool in developing this work, including for exploring conjectures, deriving and checking mathematical arguments, identifying relevant literature, and assisting with language editing and typesetting. All AI-generated suggestions, calculations, proofs, and references were critically examined and independently verified by the authors. The authors made all final scientific decisions and take full responsibility for the accuracy, originality, and content of the manuscript.

\vspace{0.2cm}

\emph{Note.---}A related work~\cite{li2026asymptoticentanglementhidingstabilizer}, which appeared approximately one week ago, studies entanglement distillation under LSCC.

\bibliography{bib.bib}

@article{vidal1999robustness,
  title = {Robustness of entanglement},
  author = {Vidal, Guifr{\'e} and Tarrach, Rolf},
  journal = {Physical Review A},
  volume = {59},
  number = {1},
  pages = {141--155},
  year = {1999},
  publisher = {American Physical Society},
  doi = {10.1103/PhysRevA.59.141}
}

@article{takagi2019general,
  title = {General Resource Theories in Quantum Mechanics and Beyond:
           Operational Characterization via Discrimination Tasks},
  author = {Takagi, Ryuji and Regula, Bartosz},
  journal = {Physical Review X},
  volume = {9},
  number = {3},
  pages = {031053},
  year = {2019},
  publisher = {American Physical Society},
  doi = {10.1103/PhysRevX.9.031053}
}

@misc{li2026asymptoticentanglementhidingstabilizer,
      title={Asymptotic Entanglement Hiding under Stabilizer Restrictions}, 
      author={Jicun Li and Wei Xie and Jun Wu and Honglin Chen and Xiang-Yang Li},
      year={2026},
      eprint={2608.18440},
      archivePrefix={arXiv},
      primaryClass={quant-ph},
      url={https://arxiv.org/abs/2608.18440}, 
}

@article{Awschalom2021,
  title = {Development of Quantum Interconnects (QuICs) for Next-Generation Information Technologies},
  author = {Awschalom, David and Berggren, Karl K. and Bernien, Hannes and Bhave, Sunil and Carr, Lincoln D. and Davids, Paul and Economou, Sophia E. and Englund, Dirk and Faraon, Andrei and Fejer, Martin and Guha, Saikat and Gustafsson, Martin V. and Hu, Evelyn and Jiang, Liang and Kim, Jungsang and Korzh, Boris and Kumar, Prem and Kwiat, Paul G. and Lon\v{c}ar, Marko and Lukin, Mikhail D. and Miller, David A. B. and Monroe, Christopher and Nam, Sae Woo and Narang, Prineha and Orcutt, Jason S. and Raymer, Michael G. and Safavi-Naeini, Amir H. and Spiropulu, Maria and Srinivasan, Kartik and Sun, Shuo and Vu\v{c}kovi\'{c}, Jelena and Waks, Edo and Walsworth, Ronald and Weiner, Andrew M. and Zhang, Zheshen},
  journal = {PRX Quantum},
  volume = {2},
  number = {1},
  pages = {017002},
  year = {2021},
  publisher = {APS},
  doi = {10.1103/PRXQuantum.2.017002}
}

@article{Briegel1998,
  title = {Quantum Repeaters: The Role of Imperfect Local Operations in Quantum Communication},
  author = {Briegel, H.-J. and D{\"u}r, W. and Cirac, J. I. and Zoller, P.},
  journal = {Physical Review Letters},
  volume = {81},
  number = {26},
  pages = {5932--5935},
  year = {1998},
  publisher = {APS},
  doi = {10.1103/PhysRevLett.81.5932}
}

@article{Monroe2014,
  title = {Large-scale modular quantum-computer architecture with atomic memory and photonic interconnects},
  author = {Monroe, C. and Raussendorf, R. and Ruthven, A. and Brown, K. R. and Maunz, P. and Duan, L.-M. and Kim, J.},
  journal = {Physical Review A},
  volume = {89},
  number = {2},
  pages = {022317},
  year = {2014},
  publisher = {APS},
  doi = {10.1103/PhysRevA.89.022317}
}

@article{Nickerson2014,
  title = {Freely Scalable Quantum Technologies Using Cells of 5-to-50 Qubits with Very Lossy and Noisy Photonic Links},
  author = {Nickerson, Naomi H. and Fitzsimons, Joseph F. and Benjamin, Simon C.},
  journal = {Physical Review X},
  volume = {4},
  number = {4},
  pages = {041041},
  year = {2014},
  publisher = {APS},
  doi = {10.1103/PhysRevX.4.041041}
}

@article{Kimble2008,
  title = {The quantum internet},
  author = {Kimble, H. J.},
  journal = {Nature},
  volume = {453},
  number = {7198},
  pages = {1023--1030},
  year = {2008},
  publisher = {Nature Publishing Group},
  doi = {10.1038/nature07127}
}

@article{Wehner2018,
  title = {Quantum internet: A vision for the road ahead},
  author = {Wehner, Stephanie and Elkouss, David and Hanson, Ronald},
  journal = {Science},
  volume = {362},
  number = {6412},
  pages = {eaam9288},
  year = {2018},
  publisher = {AAAS},
  doi = {10.1126/science.aam9288}
}

@article{Cuomo2020,
  title = {Towards a distributed quantum computing ecosystem},
  author = {Cuomo, Daniele and Caleffi, Marcello and Cacciapuoti, Angela Sara},
  journal = {IET Quantum Communication},
  volume = {1},
  number = {1},
  pages = {3--8},
  year = {2020},
  publisher = {Wiley},
  doi = {10.1049/iet-qtc.2020.0002}
}

@book{VanMeter2014,
  title = {Quantum Networking},
  author = {Van Meter, Rodney},
  publisher = {Wiley-ISTE},
  year = {2014},
  doi = {10.1002/9781118648919}
}

@article{nickerson2013topological,
  title = {Topological quantum computing with a network of imperfect qubits},
  author = {Nickerson, Naomi H. and Li, Ying and Benjamin, Simon C.},
  journal = {Nature Communications},
  volume = {4},
  number = {1},
  pages = {1756},
  year = {2013},
  publisher = {Nature Publishing Group},
  doi = {10.1038/ncomms2773}
}

@article{Gottesman1999,
  title = {Demonstrating the viability of universal quantum computation using teleportation and single-qubit operations},
  author = {Gottesman, Daniel and Chuang, Isaac L.},
  journal = {Nature},
  volume = {402},
  number = {6760},
  pages = {390--393},
  year = {1999},
  publisher = {Nature Publishing Group},
  doi = {10.1038/46503}
}

@article{Cirac1999,
  title = {Distributed quantum computation over noisy channels},
  author = {Cirac, J. I. and Ekert, A. K. and Huelga, S. F. and Macchiavello, C.},
  journal = {Physical Review A},
  volume = {59},
  number = {6},
  pages = {4249--4254},
  year = {1999},
  publisher = {APS},
  doi = {10.1103/PhysRevA.59.4249}
}

@article{Jiang2007,
  title = {Distributed quantum computation based on small quantum registers},
  author = {Jiang, Liang and Taylor, Jacob M. and S{\o}rensen, Anders S. and Lukin, Mikhail D.},
  journal = {Physical Review A},
  volume = {76},
  number = {6},
  pages = {062323},
  year = {2007},
  publisher = {APS},
  doi = {10.1103/PhysRevA.76.062323}
}

@article{Main2025,
  title = {Distributed quantum computing across an optical network link},
  author = {Main, D. and Drmota, P. and Nadlinger, D. P. and Ainley, E. M. and Agrawal, A. and Nichol, B. C. and Srinivas, R. and Araneda, G. and Lucas, D. M.},
  journal = {Nature},
  volume = {638},
  pages = {383--388},
  year = {2025},
  publisher = {Nature Publishing Group},
  doi = {10.1038/s41586-024-08404-x}
}

@article{vedral1997quantifying,
  title = {Quantifying Entanglement},
  author = {Vedral, Vlatko and Plenio, Martin B. and Rippin, M. A. and Knight, Peter L.},
  journal = {Physical Review Letters},
  volume = {78},
  number = {12},
  pages = {2275--2279},
  year = {1997},
  publisher = {APS},
  doi = {10.1103/PhysRevLett.78.2275}
}

@article{vidal2000entanglement,
  title = {Entanglement monotones},
  author = {Vidal, Guifr{\'e}},
  journal = {Journal of Modern Optics},
  volume = {47},
  number = {2-3},
  pages = {355--376},
  year = {2000},
  publisher = {Taylor \& Francis},
  doi = {10.1080/09500340008244048}
}

@article{plenio2007introduction,
  title = {An introduction to entanglement measures},
  author = {Plenio, Martin B. and Virmani, Shashank},
  journal = {Quantum Information \& Computation},
  volume = {7},
  number = {1},
  pages = {1--51},
  year = {2007},
  publisher = {Rinton Press}
}

@article{Horodecki2009,
  title = {Quantum entanglement},
  author = {Horodecki, Ryszard and Horodecki, Pawe{\l} and Horodecki, Micha{\l} and Horodecki, Karol},
  journal = {Reviews of Modern Physics},
  volume = {81},
  number = {2},
  pages = {865--942},
  year = {2009},
  publisher = {APS},
  doi = {10.1103/RevModPhys.81.865}
}

@article{Bennett1993,
  title = {Teleporting an unknown quantum state via dual classical and Einstein-Podolsky-Rosen channels},
  author = {Bennett, Charles H. and Brassard, Gilles and Cr{\'e}peau, Claude and Jozsa, Richard and Peres, Asher and Wootters, William K.},
  journal = {Physical Review Letters},
  volume = {70},
  number = {13},
  pages = {1895--1899},
  year = {1993},
  publisher = {APS},
  doi = {10.1103/PhysRevLett.70.1895}
}

@article{eisert2000optimal,
  title = {Optimal nonlocal operations},
  author = {Eisert, Jens and Jacobs, Kurt and Papadopoulos, P. and Plenio, Martin B.},
  journal = {Physical Review A},
  volume = {62},
  number = {5},
  pages = {052317},
  year = {2000},
  publisher = {APS},
  doi = {10.1103/PhysRevA.62.052317}
}

@article{Horsman2012,
  title = {Surface code quantum computing by lattice surgery},
  author = {Horsman, Clare and Fowler, Austin G. and Devitt, Simon and Van Meter, Rodney},
  journal = {New Journal of Physics},
  volume = {14},
  number = {12},
  pages = {123011},
  year = {2012},
  publisher = {IOP Publishing},
  doi = {10.1088/1367-2630/14/12/123011}
}

@article{Fowler2012,
  title = {Surface codes: Towards practical large-scale quantum computation},
  author = {Fowler, Austin G. and Mariantoni, Matteo and Martinis, John M. and Cleland, Andrew N.},
  journal = {Physical Review A},
  volume = {86},
  number = {3},
  pages = {032324},
  year = {2012},
  publisher = {APS},
  doi = {10.1103/PhysRevA.86.032324}
}

@article{litinski2019game,
  title = {A Game of Surface Codes: Large-Scale Quantum Computing with Lattice Surgery},
  author = {Litinski, Daniel},
  journal = {Quantum},
  volume = {3},
  pages = {128},
  year = {2019},
  publisher = {Verein zur F{\"o}rderung des Open Access Publizierens in den Quantenwissenschaften},
  doi = {10.22331/q-2019-03-05-128}
}

@inproceedings{Gottesman1998,
  title = {The Heisenberg Representation of Quantum Computers},
  author = {Gottesman, Daniel},
  booktitle = {Proceedings of the XXII International Colloquium on Group Theoretical Methods in Physics},
  pages = {32--45},
  year = {1998},
  publisher = {International Press},
  doi = {10.48550/arXiv.quant-ph/9807006}
}

@article{bravyi2005universal,
  title = {Universal quantum computation with ideal Clifford gates and noisy ancillas},
  author = {Bravyi, Sergey and Kitaev, Alexei},
  journal = {Phys. Rev. A},
  volume = {71},
  issue = {2},
  pages = {022316},
  numpages = {14},
  year = {2005},
  month = {Feb},
  publisher = {American Physical Society},
  doi = {10.1103/PhysRevA.71.022316},
  url = {https://link.aps.org/doi/10.1103/PhysRevA.71.022316}
}

@article{Veitch2014,
  title = {The resource theory of stabilizer quantum computation},
  author = {Veitch, Victor and Mousavian, Syed Asad Hamed and Gottesman, Daniel and Emerson, Joseph},
  journal = {New Journal of Physics},
  volume = {16},
  number = {1},
  pages = {013009},
  year = {2014},
  publisher = {IOP Publishing},
  doi = {10.1088/1367-2630/16/1/013009}
}

@article{HowardCampbell2017,
  title = {Application of a Resource Theory for Magic States to Fault-Tolerant Quantum Computing},
  author = {Howard, Mark and Campbell, Earl},
  journal = {Physical Review Letters},
  volume = {118},
  number = {9},
  pages = {090501},
  year = {2017},
  publisher = {APS},
  doi = {10.1103/PhysRevLett.118.090501}
}

@article{leone2022stabilizer,
  title = {Stabilizer R{\'e}nyi Entropy},
  author = {Leone, Lorenzo and Oliviero, Salvatore F. E. and Hamma, Alioscia},
  journal = {Physical Review Letters},
  volume = {128},
  number = {5},
  pages = {050402},
  year = {2022},
  publisher = {APS},
  doi = {10.1103/PhysRevLett.128.050402}
}

@article{Chitambar2019,
  title = {Quantum resource theories},
  author = {Chitambar, Eric and Gour, Gilad},
  journal = {Reviews of Modern Physics},
  volume = {91},
  number = {2},
  pages = {025001},
  year = {2019},
  publisher = {APS},
  doi = {10.1103/RevModPhys.91.025001}
}

@article{BravyiSmithSmolin2016,
  title = {Trading Classical and Quantum Computational Resources},
  author = {Bravyi, Sergey and Smith, Graeme and Smolin, John A.},
  journal = {Physical Review X},
  volume = {6},
  number = {2},
  pages = {021043},
  year = {2016},
  publisher = {APS},
  doi = {10.1103/PhysRevX.6.021043}
}

@article{Beverland2020,
doi = {10.1088/2058-9565/ab8963},
url = {https://doi.org/10.1088/2058-9565/ab8963},
year = {2020},
month = {may},
publisher = {IOP Publishing},
volume = {5},
number = {3},
pages = {035009},
author = {Beverland, Michael and Campbell, Earl and Howard, Mark and Kliuchnikov, Vadym},
title = {Lower bounds on the non-Clifford resources for quantum computations},
journal = {Quantum Science and Technology}
}

@article{Beverland2022,
  title = {Cost of Universality: A Comparative Study of the Overhead of State Distillation and Code Switching with Color Codes},
  author = {Beverland, Michael E. and Kubica, Aleksander and Svore, Krysta M.},
  journal = {PRX Quantum},
  volume = {2},
  issue = {2},
  pages = {020341},
  numpages = {46},
  year = {2021},
  month = {Jun},
  publisher = {American Physical Society},
  doi = {10.1103/PRXQuantum.2.020341},
  url = {https://link.aps.org/doi/10.1103/PRXQuantum.2.020341}
}

@article{Knill1998,
  title = {Power of One Bit of Quantum Information},
  author = {Knill, Emanuel and Laflamme, Raymond},
  journal = {Physical Review Letters},
  volume = {81},
  number = {25},
  pages = {5672--5675},
  year = {1998},
  publisher = {APS},
  doi = {10.1103/PhysRevLett.81.5672}
}

@article{Ollivier2001,
  title = {Quantum Discord: A Measure of the Quantumness of Correlations},
  author = {Ollivier, Harold and Zurek, Wojciech H.},
  journal = {Physical Review Letters},
  volume = {88},
  number = {1},
  pages = {017901},
  year = {2001},
  publisher = {APS},
  doi = {10.1103/PhysRevLett.88.017901}
}

@article{Cubitt2003,
  title = {Separable States Can Be Used to Distribute Entanglement},
  author = {Cubitt, Toby S. and Verstraete, Frank and D{\"u}r, Wolfgang and Cirac, J. Ignacio},
  journal = {Physical Review Letters},
  volume = {91},
  number = {3},
  pages = {037902},
  year = {2003},
  publisher = {APS},
  doi = {10.1103/PhysRevLett.91.037902}
}

@article{Bravyi2019,
  title = {Simulation of quantum circuits by low-rank stabilizer decompositions},
  author = {Bravyi, Sergey and Browne, Dan and Calpin, Padraic and Campbell, Earl and Gosset, David and Howard, Mark},
  journal = {Quantum},
  volume = {3},
  pages = {181},
  year = {2019},
  publisher = {Verein zur F{\"o}rderung des Open Access Publizierens in den Quantenwissenschaften},
  doi = {10.22331/q-2019-09-02-181}
}

@misc{Ganardi2026,
  author        = {Ganardi, Ray and Son, Jeongrak and Czartowski, Jakub and Lie, Seok Hyung and Ng, Nelly H. Y.},
  title         = {Manipulating heterogeneous quantum resources over a network},
  year          = {2026},
  eprint        = {2602.17803},
  archivePrefix = {arXiv},
  primaryClass  = {quant-ph},
  url           = {https://arxiv.org/abs/2602.17803}
}

@article{FangLiu2019,
 title = {No-Go Theorems for Quantum Resource Purification},
  author = {Fang, Kun and Liu, Zi-Wen},
  journal = {Phys. Rev. Lett.},
  volume = {125},
  issue = {6},
  pages = {060405},
  numpages = {7},
  year = {2020},
  month = {Aug},
  publisher = {American Physical Society},
  doi = {10.1103/PhysRevLett.125.060405},
  url = {https://link.aps.org/doi/10.1103/PhysRevLett.125.060405}
}

@article{ding2022no,
  title={No-go theorems for deterministic purification and probabilistic enhancement of coherence},
  author={Ding, Qiming and Liu, Quancheng},
  journal={Journal of Physics A: Mathematical and Theoretical},
  volume={55},
  number={10},
  pages={105301},
  year={2022},
  publisher={IOP Publishing},
  doi={https://doi.org/10.1088/1751-8121/ac4ecd}
}

@article{FangLiu2022,
  title = {No-Go Theorems for Quantum Resource Purification: New Approach and Channel Theory},
  author = {Fang, Kun and Liu, Zi-Wen},
  journal = {PRX Quantum},
  volume = {3},
  issue = {1},
  pages = {010337},
  numpages = {19},
  year = {2022},
  month = {Mar},
  publisher = {American Physical Society},
  doi = {10.1103/PRXQuantum.3.010337},
  url = {https://link.aps.org/doi/10.1103/PRXQuantum.3.010337}
}

@article{Peres1996,
  title = {Separability Criterion for Density Matrices},
  author = {Peres, A.},
  journal = {Phys. Rev. Lett.},
  volume = {77},
  pages = {1413--1415},
  year = {1996},
  doi = {10.1103/PhysRevLett.77.1413}
}

@article{Horodecki1997,
  title = {Separability of mixed states: necessary and sufficient conditions},
  author = {Horodecki, P.},
  journal = {Phys. Lett. A},
  volume = {232},
  pages = {333--339},
  year = {1997},
  doi = {10.1016/S0375-9601(97)00416-7}
}

@article{HouCaoYang2026,
  author        = {Hou, Zong-Yue and Cao, ChunJun and Yang, Zhi-Cheng},
  title         = {Stabilizer entanglement enhances magic injection},
  journal       = {{npj Quantum Information}},
  volume        = {12},
  pages         = {113},
  year          = {2026},
  doi           = {10.1038/s41534-026-01265-4},
}

@article{Elitzur1992,
title = {Quantum nonlocality for each pair in an ensemble},
journal = {Physics Letters A},
volume = {162},
number = {1},
pages = {25-28},
year = {1992},
issn = {0375-9601},
doi = {https://doi.org/10.1016/0375-9601(92)90952-I},
url = {https://www.sciencedirect.com/science/article/pii/037596019290952I},
author = {Avshalom C. Elitzur and Sandu Popescu and Daniel Rohrlich}
}

@article{Lewenstein1998,
  title = {Separability and Entanglement of Composite Quantum Systems},
  author = {Lewenstein, M. and Sanpera, A.},
  journal = {Phys. Rev. Lett.},
  volume = {80},
  issue = {10},
  pages = {2261--2264},
  year = {1998},
  month = {Mar},
  publisher = {American Physical Society},
  doi = {10.1103/PhysRevLett.80.2261},
  url = {https://link.aps.org/doi/10.1103/PhysRevLett.80.2261}
}

@article{Bu2018,
  title = {Asymmetry and coherence weight of quantum states},
  author = {Bu, K. and Anand, N. and Singh, U.},
  journal = {Phys. Rev. A},
  volume = {97},
  issue = {3},
  pages = {032342},
  numpages = {10},
  year = {2018},
  month = {Mar},
  publisher = {American Physical Society},
  doi = {10.1103/PhysRevA.97.032342},
  url = {https://link.aps.org/doi/10.1103/PhysRevA.97.032342}
}

@article{Yao2020,
  title = {Anomalies of the weight-based coherence measure and mixed maximally coherent states},
  author = {Yao, Y. and Li, D. and Sun, C. P.},
  journal = {Phys. Rev. A},
  volume = {102},
  issue = {3},
  pages = {032406},
  numpages = {8},
  year = {2020},
  month = {Sep},
  publisher = {American Physical Society},
  doi = {10.1103/PhysRevA.102.032406},
  url = {https://link.aps.org/doi/10.1103/PhysRevA.102.032406}
}

@article{Ducuara2020,
  title = {Operational Interpretation of Weight-Based Resource Quantifiers in Convex Quantum Resource Theories},
  author = {Ducuara, A. F. and Skrzypczyk, P.},
  journal = {Phys. Rev. Lett.},
  volume = {125},
  issue = {11},
  pages = {110401},
  numpages = {6},
  year = {2020},
  month = {Sep},
  publisher = {American Physical Society},
  doi = {10.1103/PhysRevLett.125.110401},
  url = {https://link.aps.org/doi/10.1103/PhysRevLett.125.110401}
}

@article{Uola2020,
  title = {All Quantum Resources Provide an Advantage in Exclusion Tasks},
  author = {Uola, R. and Bullock, T. and Kraft, T. and Pellonp{\"a}{\"a}, J.-P. and Brunner, N.},
  journal = {Phys. Rev. Lett.},
  volume = {125},
  issue = {11},
  pages = {110402},
  numpages = {6},
  year = {2020},
  month = {Sep},
  publisher = {American Physical Society},
  doi = {10.1103/PhysRevLett.125.110402},
  url = {https://link.aps.org/doi/10.1103/PhysRevLett.125.110402}
}

@book{CLRS,
  author    = {Thomas H. Cormen and Charles E. Leiserson
               and Ronald L. Rivest and Clifford Stein},
  title     = {Introduction to Algorithms},
  edition   = {3},
  publisher = {MIT Press},
  address   = {Cambridge, MA},
  year      = {2009},
  isbn      = {978-0-262-03384-8}
}

@article{Heimendahl_2022,
   title={The axiomatic and the operational approaches to resource theories of magic do not coincide},
   volume={63},
   ISSN={1089-7658},
   url={http://dx.doi.org/10.1063/5.0085774},
   DOI={10.1063/5.0085774},
   number={11},
   journal={Journal of Mathematical Physics},
   publisher={AIP Publishing},
   author={Heimendahl, Arne and Heinrich, Markus and Gross, David},
   year={2022},
   month=Nov }

\clearpage
\onecolumngrid

\begin{center}
{\large\bfseries Supplemental Material for\\[2pt]
``Distributed Resource Theory of Entanglement and Magic''}
\end{center}

\vspace{6pt}

\setcounter{section}{0}
\setcounter{equation}{0}
\setcounter{figure}{0}
\setcounter{table}{0}
\setcounter{theorem}{0}

\setcounter{secnumdepth}{3}

\setcounter{section}{0}
\renewcommand{\thesection}{S\Roman{section}}

\setcounter{equation}{0}
\renewcommand{\theequation}{S\arabic{equation}}

\setcounter{figure}{0}
\renewcommand{\thefigure}{S\arabic{figure}}

\setcounter{table}{0}
\renewcommand{\thetable}{S\arabic{table}}

\setcounter{theorem}{0}
\renewcommand{\thetheorem}{S\arabic{theorem}}


Here, we elaborate the details and proofs for results in the main text.

\section{Operational DREAM framework}
\label{sec:sm-framework}
We first introduce the basic framework of the distributed resource theory of entanglement and magic (DREAM). DREAM is motivated by distributed fault-tolerant quantum computing, in which both entanglement and magic constitute essential resources. Conventional approaches generally treat these two resources separately: magic enables non-Clifford logical operations, whereas entanglement enables nonlocal quantum operations. In contrast, we show that they admit a unified and coherent resource-theoretic framework that captures both resources simultaneously.

\subsection{Fault-tolerant quantum network model}

We consider a quantum network described by a finite graph $G=(V,E)$, where each vertex $v\in V$ represents a local quantum system controlling a finite collection of logical qubits, and each edge $e\in E$ represents a quantum channel connecting two local systems.

Entanglement and magic capture two distinct quantum resources in this setting. Entanglement characterizes nonlocal quantum correlations shared among local systems, whereas magic characterizes the nonstabilizer resource required to implement operations beyond the stabilizer formalism. To treat these resources within a unified framework, we adopt the resource-theoretic perspective and first specify the corresponding free operations. In the resource theories of entanglement and magic, the free operations are local operations and classical communication (LOCC) and stabilizer operations, respectively. This naturally motivates us to define the free operations for the joint distributed setting as local stabilizer operations and classical communication (LSCC). 
Specifically, an LSCC protocol is a finite sequence of local stabilizer operations, interspersed with classical communication and assisted by shared randomness. At each node, the allowed local operations are generated by (i) preparation of pure or mixed stabilizer ancillas, (ii) Clifford unitaries or isometries, (iii) Pauli measurements with classical feed-forward, and (iv) discarding of subsystems.

By construction, LSCC cannot generate either nonlocal entanglement or local nonstabilizer resources from free inputs. Consequently, logical Bell pairs shared between different nodes, local nonstabilizer states, and, more generally, any shared state that cannot be prepared by LSCC must be regarded as nontrivial resources.

According to the golden rule of resource theories, the set of free resource under DREAM is defined as 
\begin{equation}
 \F_V=\operatorname{conv}\left\{\bigotimes_{v\in V}\sigma_v:\sigma_v\in\STAB_v\right\},
 \label{eq:Sfree}
\end{equation}
where $\STAB_v$ is the stabilizer polytope on node $v$.
Equivalently, one may restrict the extreme elements to pure stabilizer states.  Shared randomness prepares every state in Eq.~\eqref{eq:Sfree}; conversely LSCC maps it into itself.  It is useful to keep three larger sets distinct:
\begin{equation}
 \F_V\subseteq\SEP_V,\qquad \F_V\subseteq\STAB_{\rm global},
 \qquad \F_V\subsetneq\SEP_V\cap\STAB_{\rm global}.
 \label{eq:setinclusions}
\end{equation}
The strict inclusion is proved constructively below. We show that it gives rise to a resource structure that is much richer than a simple combination of entanglement and magic.

The operational class is also smaller than the full set of resource-nongenerating maps
\begin{equation}
 \operatorname{RNG}(\F_V)=\{\Lambda:\Lambda(\F_V)\subseteq\F_V\}.
\end{equation}
For example, a nonlocal SWAP between equal-dimensional nodes preserves the free set but transmits an unknown state and is not LSCC.  Results concerning communication, Kraus branches, or Bell cost below use operational LSCC, never consider this maximal relaxation.  The distinction mirrors the known gap between operational and axiomatic stabilizer operations~\cite{Heimendahl_2022}.

\subsection{Convex measures}

For a closed convex free set $\F$, the generalized and standard robustness are defined as 
\begin{align}
 R_{\rm g}^{\F}(\rho)&=\min\left\{s\geqslant0:\frac{\rho+s\omega}{1+s}\in\F,
 \ \omega\ \text{a state}\right\},\label{eq:Rgdef}\\
 R_{\rm s}^{\F}(\rho)&=\min\left\{s\geqslant0:\frac{\rho+s\sigma_-}{1+s}=\sigma_+,
 \ \sigma_\pm\in\F\right\}.
 \label{eq:Rsdef}
\end{align}
The free base norm is
\begin{equation}
 \Gamma_{\F}(\rho)=\min\left\{\sum_j|q_j|:\rho=\sum_jq_j\sigma_j,
 \ \sigma_j\in\F,\ \sum_jq_j=1\right\}
 =1+2R_{\rm s}^{\F}(\rho).
 \label{eq:gammadef}
\end{equation}
The dual forms used below are
\begin{align}
 1+R_{\rm g}^{\F}(\rho)&=\max_{W\ge0}\{\Tr(W\rho):
 \Tr(W\sigma)\leqslant1\ \forall\sigma\in\F\},\label{eq:Rgdual}\\
 \Gamma_{\F}(\rho)&=\max_H\{\Tr(H\rho):
 |\Tr(H\sigma)|\leqslant1\ \forall\sigma\in\F\}.
 \label{eq:gammadual}
\end{align}
All are convex, faithful, and nonincreasing under LSCC for $\mathcal F = \mathcal F_V$.  We also consider the maximal free weight
\begin{equation}
 w_{\F}(\rho)=\max\{\lambda:\rho=\lambda\sigma+(1-\lambda)\omega,
 \ \sigma\in\F,\ \omega\ \text{being an arbitrary state}\}.
 \label{eq:freeweight}
\end{equation}

For a pure $n$-qubit state, its Pauli stabilizer group is
\begin{equation}
 \operatorname{Stab}(\psi)=\{P\in\mathcal P_n:P\ket\psi=\ket\psi\},
\end{equation}
and its stabilizer nullity \cite{Beverland2020} is defined as
\begin{equation}
 \nu(\psi)=n-\log_2|\operatorname{Stab}(\psi)|.
 \label{eq:Snullity}
\end{equation}
It is additive on pure tensor products and invariant under Clifford unitaries. It can also be extended to mixed states via the convex roof construction,
\begin{equation}
 \nu(\rho)=\min_{\rho=\sum_jp_j\ketbra{\psi_j}}\sum_jp_j\nu(\psi_j).
 \label{eq:Snullitymixed}
\end{equation}
The stabilizer nullity of a mixed state is always upper bounded by the number of qubits, i.e. $\nu(\rho)\leqslant n$ for any $n$-qubit state $\rho$.

\section{Separable global-stabilizer states with nonzero DREAM resource}

\label{sec:sep_stab_resource}

The free set of DREAM is the convex hull of all product stabilizer states across the node partition. A natural implication is that any state in the free set is separable and magic-free. However, the converse is not true: not every separable magic-free state is free in the sense of DREAM. In this section, we construct an explicit two-node family of such states and prove their properties.

We define the single-qubit measurement operator $M=(X+Y)/\sqrt2$ and its $\pm 1$ eigenstates $\tau_\pm=(\id\pm M)/2$. The two-node family is defined as
\begin{equation}
  \rho_\eta=\frac14(\id+\eta M\otimes M),
  \label{eq:Srhofamily}
\end{equation}
where $|\eta|\le1$ ensures positivity of $\rho_\eta$. We have the following proposition, which characterizes the geometry of this family.
\begin{proposition}[Exact geometry of the two-node family]
For every $\rho_{\eta}$ given in \eqref{eq:Srhofamily}, we have the following properties:
\begin{enumerate}
\renewcommand{\labelenumi}{(\arabic{enumi})}
  \item $\rho_\eta\in\SEP_{AB}\cap\STAB_{AB}$ for all $|\eta|\leqslant1$;
  \item $\rho_\eta\in\F_{AB}$ if and only if $|\eta|\leqslant\tfrac12$;
  \item The robustness measures and base norm are given by
  \begin{equation}
      \begin{aligned}
   R_{\rm g}^{\F}(\rho_\eta)&=\max\left\{0,\frac{2|\eta|-1}{3}\right\},\nonumber\\
   R_{\rm s}^{\F}(\rho_\eta)&=\max\left\{0,|\eta|-\frac12\right\},\nonumber\\
   \Gamma_{\F}(\rho_\eta)&=\max\left\{1,2|\eta|\right\}.
  \end{aligned}
  \end{equation}

\end{enumerate}
\label{prop:Sgeometry}
\end{proposition}
Particularly, one of the most resourceful state of the family, $\rho_{\rm b}=\rho_1$ given by $\eta=1$, is still nondistillable by finite-round LSCC into either a Bell pair or a pure nonstabilizer state, even on a branch of nonzero probability. Consequently, the set $(\SEP\cap\STAB_{AB})\setminus\F_{AB}$ has nonzero volume. The proof of Proposition~\ref{prop:Sgeometry} is provided in the following subsections.

\subsection{Separability and global stabilizerness}
Define the boundary states $\rho_1=\rho_{\rm b}$ and $\rho_{-1}=(Z\otimes\id)\rho_{\rm b}(Z\otimes\id)$. Then any state $\rho_\eta$ can be expressed as a convex combination of maximally mixed state and a boundary state:
\begin{equation}
  \begin{aligned}
    \rho_\eta&=(1-\eta)\frac{\id}{4}+\eta\rho_{\rm b},\qquad 0\le\eta\leqslant1,\\
    \rho_\eta&=(1-|\eta|)\frac{\id}{4}+|\eta|(Z\otimes\id)\rho_{\rm b}(Z\otimes\id),\qquad -1\leqslant\eta<0.
  \end{aligned}
\end{equation}
To prove (1) of Proposition~\ref{prop:Sgeometry}, it suffices to show that $\rho_{\rm b}$ is separable and a global stabilizer. It is straightforward to verify that 
\begin{equation}
    \rho_{\rm b}=\frac12(\tau_+\otimes\tau_++\tau_-\otimes\tau_-),
\end{equation}
indicating that both $\rho_{\rm b}$ and $(Z\otimes\id)\rho_{\rm b}(Z\otimes\id)$ are separable. Meanwhile, we can also express the boundary state as a convex mixture of stabilizer states. To see this, we may expand $\rho_{\rm b}$ in the Pauli basis as
\begin{equation}
  \rho_{\rm b}=\frac14\id+\frac18(XX+XY+YX+YY) = \frac12\left[\frac14\id+\frac14(XX+YY-ZZ)\right]+\frac12\left[\frac14\id+\frac14(XY+YX+ZZ)\right],
\end{equation}
with the two terms are density matrices of pure stabilizer states $\ket{\Psi^+}=(\ket{01}+\ket{10})/\sqrt2$ and $\ket{\Phi_i}=(\ket{00}+i\ket{11})/\sqrt2$ with stabilizer $\{XX,YY,-ZZ\}$ and $\{XY,YX,ZZ\}$, respectively. Therefore, we establish that $\rho_{\rm b}\in \mathrm{STAB}_{AB}$. In all, we have shown that $\rho_\eta\in\SEP_{AB}\cap\STAB_{AB}$ for all $|\eta|\le1$.


\subsection{Threshold for DREAM-resourcefulness}
In this section, we show the threshold for $\rho_\eta$ to be resourceful in the sense of DREAM. To do so, we introduce the witness operator $J=(X+Y)\otimes(X+Y)$. For any product of single-qubit stabilizer states $\sigma_A\otimes\sigma_B$, we have
\begin{equation}
 |\langle J\rangle_{\sigma_A\otimes\sigma_B}|=|\langle X+Y\rangle_{\sigma_A}\langle X+Y\rangle_{\sigma_B}|\leqslant 1,
\end{equation}
where the inequality follows from the fact that the Bloch vector of a single-qubit stabilizer can only take values in $\{\pm e_x,\pm e_y,\pm e_z\}$. By convexity, the inequality extends to any product stabilizer mixture, namely the DREAM-free set $\F_{AB}$ by definition. That means, any $\rho_\eta$ satisfying $|\langle J\rangle_{\rho_\eta}|>1$ is resourceful. Taking expectation value of $J$ on $\rho_\eta$, we have
\begin{equation}
  \begin{aligned}
    \langle J\rangle_{\rho_\eta} &= \Tr(J\rho_\eta)\\
    &= \Tr\left[(2M\otimes M)\cdot\frac14(\id+\eta M\otimes M)\right]=2\eta,
  \end{aligned}
\end{equation}
suggesting that $\rho_\eta$ is resourceful for any $|\eta|>1/2$. Furthermore, we can show that the bound is tight. By convexity of $\F_{AB}$, it suffices to show that $\rho_{\pm 1/2}\in\F_{AB}$. Rewriting $\rho_{\pm 1/2}$ in the Pauli basis, we have
\begin{equation}
  \begin{aligned}
    \rho_{\pm 1/2} &= \frac14\left(\id\pm\frac14(X+Y)\otimes(X+Y)\right)\\
    &= \frac14\left(\id\pm\frac14(XX+XY+YX+YY)\right)\\
    &= \frac14\left(\frac14\id\pm\frac14 XX\right)+\frac14\left(\frac14\id\pm\frac14 XY\right) + \frac14\left(\frac14\id\pm\frac14 YX\right) + \frac14\left(\frac14\id\pm\frac14 YY\right),
  \end{aligned}
\end{equation}
where $\frac14\id\pm\frac14 P_iP_j$ is a mixture of product stabilizer states for $P_i,P_j\in\{X,Y\}$. Therefore, the expression above explicitly shows that $\rho_{\pm 1/2}$ is a convex mixture of product stabilizer states, and hence $\rho_{\pm 1/2}\in\F_{AB}$. In summary, we have shown that $\rho_\eta$ is free in the sense of DREAM if and only if $|\eta|\leqslant 1/2$.



\subsection{Base norm and robustness measures}

By definition, the base norm of $\rho_\eta$ equals $1$ for all free states with $|\eta|\leqslant 1/2$. For $|\eta|> 1/2$, we could expand $\rho_\eta$ in a DREAM-free basis as
\begin{equation}
  \rho_\eta=\left(\eta+\frac12\right)\rho_{1/2}-\left(\eta-\frac12\right)\rho_{-1/2}.
\label{eq:Ssigned}
\end{equation}
This feasible free decomposition gives an upper bound of the base norm of $\rho_\eta$, i.e. $\Gamma(\rho_\eta)\leqslant|\eta+1/2|+|\eta-1/2|= 2|\eta|$. To show that this value is also a lower bound, we resort to the dual program defined by Eq.~\eqref{eq:gammadual}, where we take the trial operator $H=\mathrm{sgn}(\eta)J$ with $J=(X+Y)\otimes(X+Y)$ defined earlier, satisfying $|\Tr(\mathrm{sgn}(\eta)J\sigma)|\leqslant1$ for all $\sigma\in\mathcal{F}_{\{A,B\}}$. Then $\Gamma(\rho_\eta)\geqslant \mathrm{sgn}(\eta)\langle J\rangle_{\rho_\eta}=2|\eta|$ provides a lower bound of the base norm. Therefore, we conclude $\Gamma(\rho_\eta)=2|\eta|$ for $|\eta|> 1/2$. Combining with the free cases $|\eta|\leqslant 1/2$ yields the base norm result given in part (3) of Proposition~\ref{prop:Sgeometry},
\begin{equation}
  {\ \Gamma_{\F}(\rho_\eta)=\max\{1,2|\eta|\}.\ }
\end{equation}
The relationship Eq.~\eqref{eq:gammadef} between the base norm and the standard robustness directly leads to  
\begin{equation}
  {\ R_{\rm s}^{\F}(\rho_\eta)=\max\{0,|\eta|-\tfrac12\}.\ }
\end{equation}
To investigate the generalized robustness of $\rho_\eta$, we need to find the least value of $s$ such that for some free state $\sigma\in\mathcal{F}_{\{A,B\}}$ we have $(1+s)\sigma-\rho_\eta\geqslant 0$. For $\eta>1/2$, we take $\rho_{1/2}$ as the free state, and the above expression writes
\begin{equation}
  \begin{aligned}
    (1+s)\rho_{1/2}-\rho_\eta &= (1+s)\frac14(\id+\frac12 M\otimes M) - \frac14(\id+\eta M\otimes M)\\
    &= \frac14(s\id + \frac{1+s-2\eta}{2} M\otimes M).
  \end{aligned}
\end{equation}
Then positivity dictates $s\geqslant |1+s-2\eta|/2$, or equivalently $s\geqslant (2\eta-1)/3$. Hence $(2\eta-1)/3$ is an upper bound of the generalized robustness $R_{\rm g}^{\F}(\rho_\eta)$ for $\eta>1/2$. For the case $\eta<-1/2$, taking $\rho_{-1/2}$ as the free state leads to similar argument, which provides the upper bound $R_{\rm g}^{\F}(\rho_\eta)\leqslant (2|\eta|-1)/3$. To establish equality, we take the normalized positive operator $W = (2\id + \mathrm{sgn}(\eta)J)/3$, which satisfies $\Tr(W\sigma)\leqslant 1$ for all free states $\sigma\in\F_{AB}$ and apply the dual program Eq.~\eqref{eq:Rgdual}. This leads to a lower bound for $R_{\rm g}^{\F}(\rho_\eta)$:
\begin{equation}
    1+R_{\rm g}^{\F}(\rho_\eta)\geqslant \Tr(W\rho_\eta)=\frac{2+2|\eta|}{3},
\end{equation}
or equivalently, $R_{\rm g}^{\F}(\rho_\eta)\geqslant (2|\eta|-1)/3$. Combining with the results for the free states, we have 
\begin{equation}
 {\ R_{\rm g}^{\F}(\rho_\eta)
 =\max\left\{0,\frac{2|\eta|-1}{3}\right\},\ }
\end{equation}
validating the generalized robustness result given in part (3) of Proposition~\ref{prop:Sgeometry}.

\section{Multipartite bound correlations}
\label{sec:sm-multipartite-bound}

The two-node example naturally extends to a multipartite system with an even number of nodes, which is also resourceful with respect to DREAM while possessing neither entanglement nor global magic. Suppose the globol system is a quantum network composed of $2k$ local qubits, then we may define the family of multipartite global states as
\begin{equation}
 \Theta_{2k}(t)=\frac1{2^{2k}}\left(I+tM^{\otimes2k}\right),\qquad |t|\leqslant1.
 \label{eq:Stheta}
\end{equation}
We now provide the following theorem as the multipartite generalization of Proposition~\ref{prop:Sgeometry}, characterizing the geometry of the family $\Theta_{2k}(t)$ and its resourcefulness with respect to DREAM.



\begin{theorem}[Exact geometry of the multipartite family]
For every integer $k\ge1$ and $|t|\leqslant1$, the state $\Theta_{2k}(t)$ given in \eqref{eq:Stheta} has the following properties:
\begin{enumerate}
\renewcommand{\labelenumi}{(\arabic{enumi})}
  \item $\Theta_{2k}(t)\in\SEP_{1:\cdots:2k}\cap\STAB_{\rm global}$ for all $|t|\leqslant1  $
  \item $\Theta_{2k}(t)\in\F_{2k}$ if and only if $|t|\leqslant 2^{-k}$;
  \item The robustness measures and base norm are given by
  \begin{equation}
      \begin{aligned}
   R_{\rm g}^{\F_{2k}}(\Theta_{2k}(t))&=\max\left\{0,\frac{|t|-2^{-k}}{1+2^{-k}}\right\},\\
   R_{\rm s}^{\F_{2k}}(\Theta_{2k}(t))&=\max\left\{0,\frac{2^k|t|-1}{2}\right\},\\
   \Gamma_{\F_{2k}}(\Theta_{2k}(t))&=\max\{1,2^k|t|\}.
  \end{aligned}
  \end{equation}
\end{enumerate}

 \label{thm:Sthetatheorem_multipartite}
\end{theorem}

\subsection{Full separability, global stabilizer membership, and the free threshold}

To show the full separbility of the state $\Theta_n(t)$, we use the fact that $M=\tau_+-\tau_-$ and $\id_2=\tau_++\tau_-$, and expand $\Theta_n(t)$ as
\begin{equation}
  \begin{aligned}
    \Theta_n(t) &= \frac1{2^n}\left(\id+tM^{\otimes n}\right)\\
    &= \frac1{2^n}\left(\bigotimes_{j=1}^n(\tau_++\tau_-)+t\bigotimes_{j=1}^n(\tau_+-\tau_-)\right)\\
    &= \frac1{2^n}\sum_{\bm s\in\{\pm1\}^n}\left(1+t\prod_js_j\right)\bigotimes_{j=1}^n\tau_{s_j},
  \end{aligned}
\end{equation}
where we use the notation $\tau_{+1}=\tau_+$ and $\tau_{-1}=\tau_-$. Note that the coefficients $(1+t\prod_js_j)/2^n$ of each tensor product term is non-negative for $|t|\le1$, which proves that $\Theta_n(t)$ is a convex combination of product states and hence fully separable.

We then prove that for even number of parties, $n=2k$, the state $\Theta_{2k}(t)$ is a global stabilizer for all $|t|\le1$. To see this, we may recall the result given in Proposition~\ref{prop:Sgeometry} that the two-node state $\rho_{1}=(\id +M\otimes M)/4$ is in the two-qubit global stabilizer convex hull, 
\begin{equation}
  \rho_1 = \frac12\ketbra{\Psi^+}+\frac12\ketbra{\Phi_i}\in\STAB_{\rm global}.
\end{equation}
This inspires us to pair the $2k$ nodes into $k$ pairs, and write $\Theta_{2k}(t)$ first as a convex combination of tensor products of $k$ two-qubit states $\rho_{\pm 1}$, each belonging to the stabilizer convex hull. This could be seen by using
\begin{equation}
  M\otimes M = \frac12\left[(\id + M\otimes M) - (\id - M\otimes M)\right] = 2\left(\rho_1 - \rho_{-1}\right), \quad \id_2\otimes \id_2 = 2\left(\rho_1 + \rho_{-1}\right).
\end{equation}
Then the state $\Theta_{2k}(t)$ can be expressed as
\begin{equation}
  \begin{aligned}
    \Theta_{2k}(t) &= \frac1{2^{2k}}\left(\id+tM^{\otimes 2k}\right)\\
    &= \frac1{2^{2k}}\left(\bigotimes_{j=1}^k \id_2^{(2j-1)}\otimes \id_2^{(2j)}+t\bigotimes_{j=1}^k M^{(2j-1)}\otimes M^{(2j)}\right)\\
    &= \frac1{2^{k}}\left[\bigotimes_{j=1}^k\left(\rho_1^{(2j-1,2j)} + \rho_{-1}^{(2j-1,2j)}\right)+t\bigotimes_{j=1}^k \left(\rho_1^{(2j-1,2j)} - \rho_{-1}^{(2j-1,2j)}\right)\right]\\
    &= \frac1{2^{k}}\sum_{\bm r\in\{\pm1\}^k}\left(1+t\prod_jr_j\right)\bigotimes_{j=1}^k\rho_{r_j}^{(2j-1,2j)},
  \end{aligned}
\end{equation}
where $\rho_{r_j}^{(2j-1,2j)}$ denotes the two-qubit state $\rho_{\pm 1}$ corresponding to the label $r_j=\pm 1$ on the two qubits indiced $2j-1$ and $2j$. Each term $\bigotimes_{j=1}^k\rho_{r_j}^{(2j-1,2j)}$ is a tensor product of $k$ two-qubit stabilizer and hence is itself a global stabilizer. The stabilizerness of the whole state $\Theta_{2k}(t)$ then follows from convextity and the fact that the coefficients $(1+t\prod_jr_j)/2^{2k}$ are non-negative for $|t|\leqslant1$.

\subsection{Threshold for DREAM-resourcefulness}

The DREAM-witness operator for $n$-node global state is given by $J_n=(X+Y)^{\otimes n}=2^{n/2}M^{\otimes n}$, which satisfies $|\langle J_n\rangle_\sigma|\leqslant 1$ for all $\sigma\in\F_n$. Taking expectation value of $J_n$ on $\Theta_n(t)$ gives
\begin{equation}
  \langle J_n\rangle_{\Theta_n(t)} = \Tr(J_n\Theta_n(t)) = 2^{n/2}t,
\end{equation}
which implies that $\Theta_n(t)$ is resourceful for any $|t|>2^{-n/2}$. For the case of even number of nodes $n=2k$, we have $|t|>2^{-k}$ as the threshold for resourcefulness. Further, we could prove the tightness of this threshold in an analogous way to the previous section. First, we can show that any $2k$-qubit state of the form
\begin{equation}
  \sigma^\pm_{\bm P} = \frac1{2^{2k}}\left(\id\pm\bigotimes_{j=1}^{2k} P_j\right)
\end{equation}
must belong to the free set $\F_{2k}$, where $P_j\in\{X,Y\}$ for all $j=1,\cdots,2k$. This is because $\sigma_{\bm P}^{\pm}$ is an equal mixture of product eigenstates of the local Pauli operators $\{P_j\}_{j=1}^{2k}$ whose local eigenvalues $\{s_j\}_{j=1}^{2k}$ have prescribed product $\prod_{j=1}^{2k}s_j=\pm1$. Therefore, $\sigma^\pm_{\bm P}\in\F_{2k}$, and we can thus take the average of all such states to obtain
\begin{equation}
    \frac1{2^{2k}}\sum_{\bm P\in\{X,Y\}^{\otimes 2k}}\sigma^\pm_{\bm P} = \Theta_{2k}(\pm 2^{-k})\in\F_{2k}.
\end{equation}
So the DREAM-free threshold $|t|\leqslant 2^{-k}$ we found is tight, establishing the result (2) in Theorem~\ref{thm:Sthetatheorem_multipartite}.

\subsection{Base norm and robustness measures for multipartite scenarios}

By definition, the base norm of $\Theta_{2k}(t)$ equals $1$
for all free states with $|t|\leqslant 2^{-k}$. For
$|t|>2^{-k}$, let $\epsilon=\operatorname{sgn}(t)$. We can
expand $\Theta_{2k}(t)$ in terms of the two DREAM-free
boundary states as
\begin{equation}
  \Theta_{2k}(t)
  =
  \frac{2^k|t|+1}{2}\Theta_{2k}(\epsilon2^{-k})
  -
  \frac{2^k|t|-1}{2}\Theta_{2k}(-\epsilon2^{-k}).
  \label{eq:Stheta-signed}
\end{equation}
This feasible free decomposition gives the upper bound
$\Gamma_{\F_{2k}}(\Theta_{2k}(t))\leqslant 2^k|t|$. To
show the matching lower bound, we use the dual program in
Eq.~\eqref{eq:gammadual} and take the trial operator
$H=\operatorname{sgn}(t)J_{2k}$. Since
$|\Tr(J_{2k}\sigma)|\leqslant1$ for every
$\sigma\in\F_{2k}$, the operator $H$ is dual feasible and
gives
$\Gamma_{\F_{2k}}(\Theta_{2k}(t))
\geqslant\operatorname{sgn}(t)
\langle J_{2k}\rangle_{\Theta_{2k}(t)}
=2^k|t|$.
Combining this result with the free region
$|t|\leqslant2^{-k}$, we obtain
\begin{equation}
  \Gamma_{\F_{2k}}(\Theta_{2k}(t))
  =
  \max\{1,2^k|t|\}.
\end{equation}
The relation in Eq.~\eqref{eq:gammadef} between the base
norm and the standard robustness then directly yields
\begin{equation}
  R_{\rm s}^{\F_{2k}}(\Theta_{2k}(t))
  =
  \max\left\{0,\frac{2^k|t|-1}{2}\right\}.
\end{equation}

To evaluate the generalized robustness, we seek the smallest
$s$ such that $(1+s)\sigma-\Theta_{2k}(t)\succeq0$ for
some $\sigma\in\F_{2k}$. For $t>2^{-k}$, we take the free
boundary state $\sigma=\Theta_{2k}(2^{-k})$, giving
\begin{equation}
  \begin{aligned}
  &(1+s)\Theta_{2k}(2^{-k})-\Theta_{2k}(t)\\
  &\quad=
  \frac1{2^{2k}}
  \left[
    s\id+
    \bigl((1+s)2^{-k}-t\bigr)M^{\otimes2k}
  \right].
  \end{aligned}
\end{equation}
Since $M^{\otimes2k}$ has eigenvalues $\pm1$, positivity
requires
$s\geqslant|(1+s)2^{-k}-t|$. The smallest feasible value is
$s=(t-2^{-k})/(1+2^{-k})$, which gives an upper bound on
$R_{\rm g}^{\F_{2k}}(\Theta_{2k}(t))$. The case
$t<-2^{-k}$ follows analogously by taking
$\Theta_{2k}(-2^{-k})$, yielding the upper bound
$(|t|-2^{-k})/(1+2^{-k})$.

To prove equality, we take the normalized positive operator
\begin{equation}
  W=
  \frac{2^k\id+\operatorname{sgn}(t)J_{2k}}
       {2^k+1}.
\end{equation}
The operator $W$ is positive semidefinite and satisfies
$\Tr(W\sigma)\leqslant1$ for every $\sigma\in\F_{2k}$.
Applying the dual program in Eq.~\eqref{eq:Rgdual} gives
\begin{equation}
  1+R_{\rm g}^{\F_{2k}}(\Theta_{2k}(t))
  \geqslant
  \Tr[W\Theta_{2k}(t)]
  =
  \frac{2^k(1+|t|)}{2^k+1}.
\end{equation}
Equivalently,
$R_{\rm g}^{\F_{2k}}(\Theta_{2k}(t))
\geqslant(|t|-2^{-k})/(1+2^{-k})$.
Combining the upper and lower bounds with the free cases
gives
\begin{equation}
  R_{\rm g}^{\F_{2k}}(\Theta_{2k}(t))
  =
  \max\left\{
    0,
    \frac{|t|-2^{-k}}{1+2^{-k}}
  \right\},
\end{equation}
which completes the proof of part~(3) of
Theorem~\ref{thm:Sthetatheorem_multipartite}.

\section{Teleportation bound for magic states}

In this section we prove Corollary~4 in the main text, which states that when Alice holds $q$ independent magic states and shares $c_{\Phi}$ Bell pairs and $h_{\rm b}$ resource $\rho_1$ with Bob, the maximum number of magic states that can be teleported to Bob equals $\min\{q, c_{\Phi}+h_{\rm b}\}$. 

Suppose Bob can obtain at most $m$ magic states in this scenario vian LSCC, we need to show $m\leqslant q$ and $m\leqslant  c_{\Phi}+h_{\rm b}$. Notice that any LSCC operation is magic non-generating, imposing the requirement that Bob cannot obtain an output state with more magic than the original system, which consists of $q$ magic states at Alice's side and magic-free resources, Bell pairs and $\rho_1$'s shared by both parties. This argument directly leads to $m\leqslant q$. 

To estabilish the other part, we introduce stabilizer nullity as another magic-monotone \cite{Beverland2020}. The stabilizer nullity $\nu$ on an $n$-qubit pure state $|\psi\rangle$ is defined as
\begin{equation}
  \nu(|\psi\rangle) = n - \log_2(|\mathrm{Stab}(|\psi\rangle)|),
\end{equation}
where $\mathrm{Stab}(|\psi\rangle)$ is the stabilizer group of $|\psi\rangle$. Basically, the nullity of a pure states characterizes the number of dimensions that are not stabilized by Pauli strings. The nullity of a general state $\rho$ is defined via the convex roof, 
\begin{equation}
  \nu(\rho):=\inf_{\rho = \{p_j,|\psi_j\rangle\}}\sum_j p_j\,\nu\!\left(\lvert\psi_j\rangle\right),
\end{equation}
which is non-increasing under Clifford operations, Pauli measurements, stabilizer state preparations and qubit discarding. Consider Bob's subsystem only, Bob can only exploit the shared resources and classical information obtained from Alice to generate magic states. We may safely assume Bob begins his operations only after he receives the classical bits sent by Alice, which collapses the system to a particular outcome branch. At this point, Bob has only $c_{\Phi}$ qubits from shared Bell pairs and $h_{\rm b}$ qubits from shared $\rho_1$. From now on, any Bob's possible operation cannot increase stabilizer nullity of his subsystem. By definition, the $c_{\Phi}+h_{\rm b}$ qubits has nullity not exceeding $c_{\Phi}+h_{\rm b}$, requiring Bob's final system has nullity at most $c_{\Phi}+h_{\rm b}$, while the nullity composed of $s$ magic states has nullity exactly $s$, validating the inequality. Therefore, we conclude $s\leqslant\min\{q, c_{\Phi}+h_{\rm b}\}$. 

Combining with the achievability argument that we may produce $\min\{q, c_{\Phi}+h_{\rm b}\}$ magic states at Bob's side using quantum teleportation and the LSCC protocol in Theorem.~2, we estabilish the equality $s=\min\{q, c_{\Phi}+h_{\rm b}\}$ claimed in Corollary~4. 

\section{Feasibility criterion of magic teleportation in networks}

In this section, we prove Corollary~5, which characterizes the feasibility of distributing magic states over a quantum network. Consider an undirected graph $G=(V,E)$. Each node $v\in V$ initially holds $s_v$ perfect magic states and requires $d_v$ output magic states. Each edge $e\in E$ contains $c_e$ Bell pairs and $h_e$ independent copies of the resource state $\rho_1$, with every resource state consumable once in either direction. In order to meet all demands, we have to teleport the redundant magic states from ``magic sources'', the nodes $v$ with $s_v>d_v$, to ``magic sinks'', the nodes $v$ with $s_v<d_v$, vian LSCC between neighboring nodes in consumption of one Bell pair or one $\rho_1$ state. 

This problem can be analyzed using a framework of network flow \cite{CLRS}. By defining $u_e=c_e+h_e$ as the total number of resource states across edge $e$, the demand feasibility problem reduces to a network flow problem in a multi-source-multi-sink network, and $u_e$ denotes the two-way capacity of edge $e$. Then the required resource allocation is feasible if and only if we can find a feasible flow in the network that meet the demand of all nodes, with flow value on each directed edge denoting the number of magic states teleported in that direction, and all flow values not exceeding the corresponding edge capacities. 

It is straightforward to show the condition in Corollary~5 is necessary. If all demands are feasible, we can build a corresponding network flow satisfying the constraints for each node and each edge. Taking any subset of nodes $R\subseteq V$, we investigate its net inflow of magic states. Denote $I(R)=\sum_{e\in\delta R}(i_e-o_e)$, here $i_e$ and $o_e$ are the number of magic states teleported into  and outside of $R$ through edge $e$, respectively. We investigate the bipartition $\{R,S\setminus R\}$, the net magic increase for the subsystem $R$ equals $I(R)$. Since all demands should be met, the final number of magic states should be no less than the sum of demands of all nodes in $R$, establishing
\begin{equation}
  \sum_{v\in R}s_v + I(R)\geqslant \sum_{v\in R}d_v.
\end{equation}
Meanwhile, the teleportation through an edge $e$ consumes one resource state, thus the total resource on one edge limits the possibly teleportated magic states. Particularly, $i_e\leqslant u_e$ for all $e\in E$. Hence we have
\begin{equation}
  \begin{aligned}
    I(R)&=\sum_{e\in\delta R}(i_e-o_e)\\
    &\leqslant \sum_{e\in\delta R}i_e\leqslant \sum_{e\in \delta R}u_e.
  \end{aligned}
\end{equation}
Combining the two inequalities yields
\begin{equation}
  \begin{aligned}
    \sum_{v\in R}d_v\leqslant \sum_{v\in R}s_v+\sum_{e\in \delta R}u_e,
  \end{aligned}
\end{equation}
or equivalently, $d(R)\leqslant s(R)+\sum_{e\in \delta R}u_e$, for all $R\subseteq V$. 

To prove the sufficiency of the condition, we could exploit the max-flow-min-cut theorem \cite{CLRS}. We may first reduce the multi-source-multi-sink problem into a single-source-single-sink problem by introducing a supersource node $S$ and a supersink node $T$, and let these two nodes connect to all local nodes. We further assume that each augmented edge connected to $S$ has capacity $u_{(S, v)}=s_v$, and each augmented edge conneted to $T$ has capacity $u_{(v,T)}=d_v$. This setting is effectively assuming all magic supplies are teleported from a super magic factory denoted by $S$, and all magic demands are later teleported to a super magic sink denoted by $T$, and we can meet all demands if and only if all the edges connected to $T$ are saturated. This requirement is equivalent to that we can find a feasible network flow with value $d(V)=\sum_{v\in V}d(v)$. The augmented graph is denoted by $G'$ and its nodes $V'=V\cup \{S,T\}$.

\begin{figure}[t]
    \centering
    \begin{tikzpicture}[
        xscale=1.05,
        yscale=1.08,
        every node/.style={font=\small},
        vertex/.style={
            circle,
            draw,
            minimum size=7mm,
            inner sep=0pt,
            fill=white
        },
        terminal/.style={
            circle,
            draw,
            thick,
            minimum size=8mm,
            inner sep=0pt,
            fill=gray!15
        },
        internal/.style={
            draw=gray!55,
            line width=0.8pt
        },
        supply/.style={
            draw=blue!75!black,
            line width=1.3pt,
            -{Latex[length=2mm]}
        },
        network/.style={
            draw=red!75!black,
            line width=1.3pt,
            {Latex[length=2mm]}-{Latex[length=2mm]}
        },
        demand/.style={
            draw=orange!85!black,
            line width=1.3pt,
            -{Latex[length=2mm]}
        }
    ]

    \node[terminal] (S) at (-4.5,0) {$S$};

    \node[vertex] (v1) at (-2.15,0.85) {$v_1$};
    \node[vertex] (v2) at (-1.35,-1.45) {$v_2$};

    \node[vertex] (v3) at (1.05,1.45) {$v_3$};
    \node[vertex] (v4) at (2.10,-0.85) {$v_4$};

    \node[terminal] (T) at (4.5,0) {$T$};

    \draw[dashed,thick]
        (0,-2.25)--(0,2.25)
        node[above=2pt] {$A\,|\,B$};

    \node at (-2.10,1.95)
        {$A=\{S\}\cup(V\setminus R)$};

    \node at (2.10,1.95)
        {$B=R\cup\{T\}$};

    \draw[internal,-{Latex[length=1.8mm]}]
        (S) --
        node[
            midway,
            above,
            fill=white,
            inner sep=1pt
        ] {$s_{v_1}$}
        (v1);

    \draw[internal,-{Latex[length=1.8mm]}]
        (S) --
        node[
            midway,
            below,
            fill=white,
            inner sep=1pt
        ] {$s_{v_2}$}
        (v2);

    \draw[internal,-{Latex[length=1.8mm]}]
        (v3) --
        node[
            midway,
            above,
            fill=white,
            inner sep=1pt
        ] {$d_{v_3}$}
        (T);

    \draw[internal,-{Latex[length=1.8mm]}]
        (v4) --
        node[
            midway,
            below,
            fill=white,
            inner sep=1pt
        ] {$d_{v_4}$}
        (T);

    \draw[internal]
        (v1) to[bend right=18] (v2);

    \draw[internal]
        (v3) to[bend right=18] (v4);

    \draw[supply]
        (S)
        .. controls (-4.35,2.65) and (-0.55,2.75) ..
        node[
            pos=0.54,
            above,
            fill=white,
            inner sep=1pt
        ] {$s_{v_3}$}
        (v3);

    \draw[supply]
        (S)
        .. controls (-4.35,-2.55) and (0.15,-2.65) ..
        node[
            pos=0.55,
            below,
            fill=white,
            inner sep=1pt
        ] {$s_{v_4}$}
        (v4);

    \draw[network]
        (v1) --
        node[
            midway,
            above,
            sloped,
            fill=white,
            inner sep=1pt
        ] {$u_{13}$}
        (v3);

    \draw[network]
        (v2) --
        node[
            midway,
            below,
            sloped,
            fill=white,
            inner sep=1pt
        ] {$u_{24}$}
        (v4);

    \draw[demand]
        (v1)
        .. controls (-0.25,0.45) and (2.85,0.20) ..
        node[
            pos=0.64,
            above,
            fill=white,
            inner sep=1pt
        ] {$d_{v_1}$}
        (T);

    \draw[demand]
        (v2)
        .. controls (0.10,-1.85) and (3.15,-1.55) ..
        node[
            pos=0.64,
            below,
            fill=white,
            inner sep=1pt
        ] {$d_{v_2}$}
        (T);

    \begin{scope}[shift={(-4.25,-3.05)}]

        \draw[supply]
            (0,0)--(0.7,0);

        \node[anchor=west] at (0.8,0)
            {$S\to R:\ s(R)$};

        \draw[network]
            (3.05,0)--(3.75,0);

        \node[anchor=west] at (3.85,0)
            {$\delta R:\ \sum_{e\in\delta R}u_e$};

        \draw[demand]
            (6.55,0)--(7.25,0);

        \node[anchor=west] at (7.35,0)
            {$(V\setminus R)\to T:\ d(V\setminus R)$};

    \end{scope}

    \end{tikzpicture}

    \caption{
        Reduction of the multiple-source multiple-sink
        magic-teleportation problem to a single-source
        single-sink network-flow problem. The original
        vertices are divided by an arbitrary $S$--$T$ cut
        into $V\setminus R$ on the supersource side and $R$
        on the supersink side. The edges crossing the cut
        consist of the supply edges from $S$ to $R$, the
        original network edges in $\delta R$, and the demand
        edges from $V\setminus R$ to $T$. The resulting cut
        capacity is
        \mbox{$s(R)+\sum_{e\in\delta R}u_e+d(V\setminus R)$}.
    }
    \label{fig:augmented-network-cut}
\end{figure}
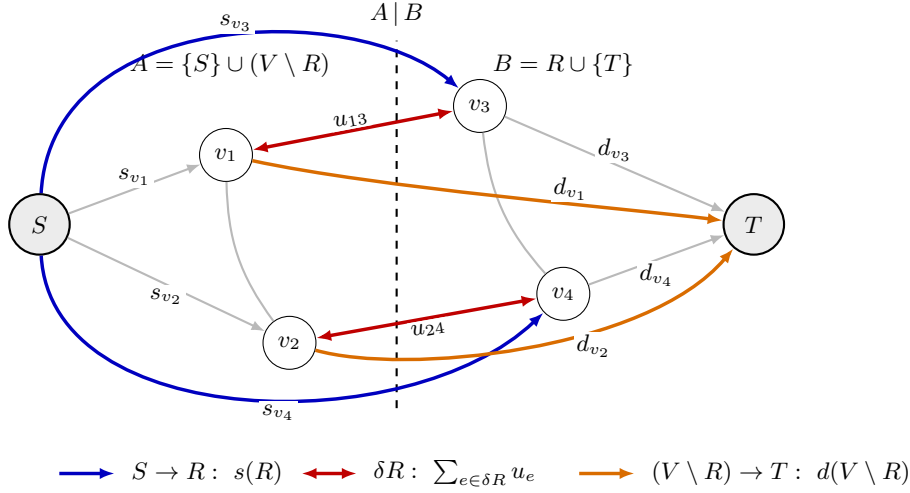

Suppose the condition in Corollary~5 is satisfied for all $R\subseteq V$ and we want to show the maximum flow value could achieve $d(V)$. According to max-flow-min-cut theorem, we only need to prove that the minimum cut has the value of $d(V)$. For an arbitrary cut $\{A|B\}$ in the augmented graph $G'$ with $S\in A$ and $T\in B$, all forward edges from $A$ to $B$ fall into three categories: (1) $(S,v')$ for $v'\in B\setminus \{T\}$, (2) $(v,v')$ for $v\in A\setminus\{S\}$ and $v'\in B\setminus \{T\}$, (3) $(v, T)$ for $v\in A\setminus\{S\}$. We now take $R=B\setminus \{T\}$, then $A\setminus\{S\}=V\setminus R$. The forward edge capacities contributed by (1) is
\begin{equation}
  \sum_{v'\in R}u_{(S,v')}=\sum_{v'\in R}s(v')=s(R).
\end{equation}
The forward edge capacities contributed by (3) is
\begin{equation}
  \sum_{v\in V\setminus R}u_{(v,T)}=\sum_{v\in V\setminus R}d(v)=d(V\setminus R).
\end{equation}
And the capacities contribute by (2) is
\begin{equation}
  \sum_{v\in V\setminus R, v'\in R}u_{(v,v')}=\sum_{e\in\delta R}u_e.
\end{equation}
A conceptual diagram is given in Fig.~\ref{fig:augmented-network-cut} to illustrate an arbitrary cut and corresponding forward edges. Summing up the three parts yields the total capacity of the cut
\begin{equation}
  C(\{A|B\})=s(R)+d(V\setminus R)+\sum_{e\in\delta R}u_e.
\end{equation}
Using the condition for the subset $R$, we have 
\begin{equation}
  \begin{aligned}
    C(\{A|B\})&=s(R)+d(V\setminus R)+\sum_{e\in\delta R}u_e\\
    &\geqslant d(R)+d(V\setminus R)\\
    &=d(V).
  \end{aligned}
\end{equation}
Therefore, the capacity of an arbitrary cut must equal $d(V)$, hence proving the feasibility of a maximum flow with value $d(V)$. To explicitly construct a magic teleportation scheme, we may apply any classical network flow algorithm such as the Ford-Fulkerson algorithm to find a maximum flow in the augmented graph $G'$, and execute magic teleportation on each edge according to the resulting flow value. 

\section{Probabilistic teleportation bound}
We adopt the definition of a new set of free states $\mathcal Q_{B|A}$ in the main text,
\begin{equation}
 \mathcal Q_{B|A}^{(\psi^a)}:=\operatorname{conv}\!\left\{
 \alpha_A\otimes\sigma_B:\
 \alpha_A\in\mathcal S_A,\ \sigma_B\in\STAB_B
 \right\},
\end{equation}
with $\mathcal S_A = \STAB_A\cup \STAB_A(\psi^a)$ with $\STAB_A(\psi^a)$ being the set of states generated by applying stabilizer operations on $\psi^a$. By definition, it is clear that $\F_{AB}\subsetneq\mathcal Q_{B|A}$, and $\mathcal Q_{B|A}$ is closed under LSCC operations. So we may regard $\mathcal Q_{B|A}$ as an enlarged new set of free states for the probabilistic teleportation scenario. Operationally, this setting implies that Alice is able to freely prepare any number of copies of magic state $\psi^a$ and send it to Bob vian LSCC, and we want to investigate the task of sending an arbitrary pure nonstabilizer state $\psi^a$ to Bob.

We characterize the finite-round LSCC protocol by a set of instruments $\{\mathcal  P_x\}_{x\in\mathcal X}$ on the global system $AB$, with each instrument $\mathcal P_x$ being a completely positive trace non-increasing map, and the sum of all instruments $\sum_{x\in\mathcal X}\mathcal P_x$ being a trace-preserving map. We further define the subchannels on Bob's side as
\begin{equation}
  \Lambda_x^{\psi}(\omega^{AB})=\Tr_A\mathcal P_x(\psi_a\otimes\omega^{AB}),
\end{equation}
where $\psi_a$ is the supplied stated prepared by Alice, and $\omega^{AB}$ is the shared resource state. The subchannel $\Lambda_x^{\psi}$ completely characterizes the probabilistic teleportation protocol when the shared resource state $\omega^{AB}$ and the LSCC outcome $x$ are given. 

A first observation is given by the following lemma:
\begin{lemma}[Free cone preservation]
  \label{lemma:free_cone_preservation}
  For any $\omega^{AB}\in\mathcal Q_{B|A}$, and any finite-round LSCC protocol with maiginal subchannels defined by $\Lambda_x^{\psi}$ on Bob's side, we have 
  \begin{equation}
  \Lambda_x^{\psi}(\omega^{AB})\in\mathrm{cone}(\STAB_B)
  \end{equation}
  for all $x\in\mathcal X$.
\end{lemma}

This lemma immediately leads to the following corollary, which states that free states are useless for teleportation of nonstabilizer states:
\begin{corollary}[Free states are useless for teleportation]
\label{coro:free_useless}
As long as $\omega^{AB}\in\mathcal Q_{B|A}$, Bob cannot obtain any nonstabilizer state via finite-round LSCC protocol, even probabilistically.
\end{corollary}

Since $\F_{AB}$ is a proper subset of $\mathcal Q_{B|A}$, the above corollary implies that any LSCC protocol that can teleport magic states with nonzero probability entails a resource state $\omega^{AB}\notin\F_{AB}$. 

The proof to Lemma~\ref{lemma:free_cone_preservation} is straightforward. By definition, we may write $\omega^{AB}\in\mathcal Q_{B|A}$ in a convex combination of product states:
\begin{equation}
  \omega^{AB}=\sum_k p_k\alpha_A^{(k)}\otimes\sigma_B^{(k)},
\end{equation}
where $\alpha_A^{(k)}\in\mathcal S_A$ and $\sigma_B^{(k)}\in\STAB_B$ for all $k$. Then the output of the subchannel $\Lambda_x^{\psi}$ can be expressed as
\begin{equation}
  \begin{aligned}
    \Lambda_x^{\psi}(\omega^{AB})&=\Tr_A\mathcal P_x(\psi_a\otimes\omega^{AB})\\
    &=\sum_k p_k\Tr_A\mathcal P_x(\psi_a\otimes\alpha_A^{(k)}\otimes\sigma_B^{(k)})\\
    &=\sum_k p_k\sigma_B^{(k)}\,\Tr_A\mathcal P_x(\psi_a\otimes\alpha_A^{(k)})\\
    &\in \mathrm{cone}(\STAB_B).
  \end{aligned}
\end{equation}
This is a natural consequence of the fact that local operations along with classical communication cannot generate magic states from free states on Bob's side. 

If we require a deterministic teleportation protocol, we should impose further requirement on the resource state $\omega^{AB}$ shared by Alice and Bob. In particular, we have the following lemma:
\begin{lemma}[Requirements for deterministic teleportation]
  \label{lemma:Require_deterministic}
If a deterministic LSCC protocol exactly relocates a pure nonstabilizer state
$\psi$ using a resource state $\omega^{AB}$, then
\begin{equation}
 w_{\F_{AB}}(\omega^{AB})=0,
 \label{eq:Szerofreeweight}
\end{equation}
and $\omega^{AB}$ cannot be full-rank.
\end{lemma}
We prove this lemma by contradiction. Suppose that $\omega^{AB}$ has a nonzero free component, i.e. $w_{\F_{AB}}(\omega)>0$. Then we may write $\omega^{AB}$ as a convex combination of a free state $f\in\F_{AB}$ and another state $\chi$,
\begin{equation}
  \omega^{AB} = \lambda f + (1-\lambda)\chi,
\end{equation}
with $\lambda>0$. Linearity of the induced subchannel $\Lambda^{\psi}$ implies that the output of the teleportation protocol can be expressed as
\begin{equation}
  \Lambda^{\psi}(\omega^{AB}) = \lambda \Lambda^{\psi}(f) + (1-\lambda)\Lambda^{\psi}(\chi).
\end{equation}
If we require that the teleportation protocol is deterministic, then $\Lambda^{\psi}(\omega^{AB})$ must equal the pure target state $\psi$. This requires each convex component on the right-hand side to equal $\psi$. However, by Lemma~\ref{lemma:free_cone_preservation}, $\Lambda^{\psi}(f)\in\mathrm{cone}(\STAB_B)$, resulting in a contradiction since $\psi\notin\mathrm{cone}(\STAB_B)$. Therefore, we conclude that $w_{\F_{AB}}(\omega^{AB})=0$. 

To prove that $\omega^{AB}$ cannot be full-rank by contradiction, we assume that $\omega^{AB}$ is full-rank. Then there exists a small $\epsilon>0$ such that the Löwner order inequality holds:
\begin{equation}
  \omega^{AB} \succcurlyeq \epsilon \id_{AB}.
\end{equation}
For any LSCC induced subchannel $\Lambda_x^{\psi}$, we have
\begin{equation}
    \Lambda_x^{\psi}(\omega^{AB}) \succcurlyeq\epsilon \Lambda_x^{\psi}(\id_{AB}).
\end{equation}
For deterministic teleportation, we require that $\Lambda_x^{\psi}(\omega^{AB})=p_x\psi$ with $p_x\geqslant0$ being the corresponding probability. This requires that $\Lambda_x^{\psi}(\id_{AB})\preccurlyeq p_x\psi$, implying that $\Lambda_x^{\psi}(\id_{AB})$ is supported only on the one-dimensional subspace spanned by $|\psi\rangle$. Meanwhile, Lemma~\ref{lemma:free_cone_preservation} implies that $\Lambda_x^{\psi}(\id_{AB})\in\mathrm{cone}(\STAB_B)$ since $\id_{AB}\in\mathcal Q_{B|A}$, meaning that $\Lambda_x^{\psi}(\id_{AB})$ can only take $0$. But this is impossible since 
\begin{equation}
  \|\omega^{AB}\|_{\infty}\Lambda_x^{\psi}(\id_{AB})\succcurlyeq\Lambda_x ^{\psi}(\omega^{AB})=p_x\psi,
\end{equation}
for some event of nonzero probability $p_x>0$. Therefore, we conclude that $\omega^{AB}$ cannot be full-rank.\\

Finally we give the robustness teleportation laws. 
\begin{theorem}[Strong directed teleportation laws]
\label{thm:Steleport}
Every finite-round LSCC instrument that teleport $\psi^a$ from Alice to Bob using resource state $\omega^{AB}$ satisfies
\begin{equation}
  \begin{aligned}
    \sum_x p_x R_{\mathrm g}^{\STAB_B}(\zeta_x)
    &\leqslant
    R_{\mathrm g}^{\mathcal Q_{B|A}^{(\psi^a)}}(\omega^{AB}),
    \label{eq:SstrongRg}\\
    \sum_x p_x R_{\mathrm s}^{\STAB_B}(\zeta_x)
    &\leqslant
    R_{\mathrm s}^{\mathcal Q_{B|A}^{(\psi^a)}}(\omega^{AB}).
  \end{aligned}
\end{equation}
\end{theorem}
In particular, if we focus on one heralded event that outputs a prescribed
state $\psi$ with probability $p$, the contribution of this event only is also bounded by the robustness of the resource state $\omega^{AB}$. Using this fact and the relation for the free sets $\F_{AB}\in $$\mathcal Q_{B|A}^{(\psi^a)}$, we have the following corollary:
\begin{corollary}[Probabilisitic teleportation bounds]
  Suppose a  probabilistic LSCC protocol exist to teleport $\psi^a$ using  $\omega^{AB}$ with probability $p$, then
\begin{equation}
  \begin{aligned}
    \label{eq:SstrongRg}
    p R_{\mathrm s}^{\STAB_B}(\psi^a)
    &\leqslant
    R_{\mathrm s}^{\mathcal Q_{B|A}^{(\psi^a)}}(\omega^{AB})
    \leqslant
    R_{\mathrm s}^{\F_{AB}}(\omega^{AB}),\\
    p R_{\mathrm g}^{\STAB_B}(\psi^a)
    &\leqslant
    R_{\mathrm g}^{\mathcal Q_{B|A}^{(\psi^a)}}(\omega^{AB})
    \leqslant
    R_{\mathrm g}^{\F_{AB}}(\omega^{AB}).
  \end{aligned}
\end{equation}
\end{corollary}

This corollary includes the standard-robustness bounds given in Theorem~6 in the main text as well as the generalized-robustness bounds. Now we are ready to prove Theorem~\ref{thm:Steleport}. We first prove the generalized-robustness inequality. Let
\begin{equation}
    r
    :=
    R_{\mathrm g}^{\mathcal Q_{B|A}^{(\psi^a)}}(\omega^{AB}).
\end{equation}
By the definition of generalized robustness, there exist
$q\in\mathcal Q_{B|A}^{(\psi)}$ and $\chi\in\mathcal D_{AB}$
such that
\begin{equation}
    \omega^{AB}=(1+r)q-r\chi,
    \label{eq:directed-generalized-decomposition}
\end{equation}
with $q\in\mathcal Q_{B|A}^{(\psi)}$ and $\chi\in\mathcal D_{AB}$. Take any finite-round LSCC instrument
$\{\mathcal P_x\}_{x\in\mathcal X}$ and the corresponding induced subchannels $\Lambda_x^\psi$, we may write its output on the resource state $\omega^{AB}$ as $\Lambda_x^\psi(\omega^{AB})=p_x\zeta_x$, where $p_x$ is the probability of outcome $x$ and $\zeta_x$ is the corresponding normalized output state. Besides, we have $\Lambda_x^\psi(q)\in \mathrm{cone}(\STAB_B)$ according to Lemma~\ref{lemma:free_cone_preservation} since $q\in\mathcal Q_{B|A}^{(\psi)}$. We may then write
\begin{equation}
    \Lambda_x^\psi(q)
    =
    u_x\sigma_x,
    \qquad
    \sigma_x\in\operatorname{STAB}_B.
\end{equation}
We also represent the output of remainder state $\chi$ as
\begin{equation}
    \Lambda_x^\psi(\chi)
    =
    v_x\chi_x,
    \qquad
    \chi_x\in\mathcal D_B.
\end{equation}
With all these ingredients, we could apply the subchannel $\Lambda_x^\psi$ to Eq.~\eqref{eq:directed-generalized-decomposition}, which gives
\begin{equation}
    p_x\zeta_x
    =
    (1+r)u_x\sigma_x-rv_x\chi_x.
\end{equation}
Taking any branch with $p_x>0$, we may divide both sides by $p_x$ to obtain
\begin{equation}
    \zeta_x=\left(1+\frac{rv_x}{p_x}\right)\sigma_x-\frac{rv_x}{p_x}\chi_x,
\end{equation}
where we have used the fact that $p_x=(1+r)u_x-rv_x$. Notice that this equation is a feasible generalized-robustness decomposition of $\zeta_x$ with respect to $\operatorname{STAB}_B$, and hence
\begin{equation}
    R_{\mathrm g}^{\operatorname{STAB}_B}(\zeta_x)
    \leqslant
    \frac{rv_x}{p_x},
\end{equation}
or equivalently,
\begin{equation}
    p_x
    R_{\mathrm g}^{\operatorname{STAB}_B}(\zeta_x)
    \leqslant  rv_x.
\end{equation}
Summing over all outcomes $x$ gives
\begin{equation}
    \sum_xp_x
    R_{\mathrm g}^{\operatorname{STAB}_B}(\zeta_x)
    \leqslant
    r
    \sum_xv_x=r=
    R_{\mathrm g}^{\mathcal Q_{B|A}^{(\psi^a)}}(\omega).
    \label{eq:average-generalized-bound}
\end{equation}
Here we have used the the trace-preserving property of $\sum_x\Lambda_x^\psi$ to conclude that $\sum_xv_x=1$. \\

We next prove the corresponding standard-robustness inequality in an analogous manner. Let
There exist
$q_+,q_-\in\mathcal Q_{B|A}^{(\psi)}$ such that
\begin{equation}
    \omega^{AB}=(1+s)q_+-sq_-.
\end{equation}
According to Lemma~\ref{lemma:free_cone_preservation}, we may write the output of the subchannel $\Lambda_x^\psi$ on $q_\pm$ as
\begin{equation}
    \Lambda_x^\psi(q_\pm)
    =
    u_x^\pm\sigma_x^\pm,
    \qquad
    \sigma_x^\pm\in\operatorname{STAB}_B,
\end{equation}
where $u_x^\pm=\operatorname{Tr}\Lambda_x^\psi(q_\pm)$ and $\sigma_x^\pm$ are normalized states in $\operatorname{STAB}_B$. Applying the induced subchannel $\Lambda_x^\psi$ to the decomposition of $\omega^{AB}$ gives
\begin{equation}
    p_x\zeta_x
    =
    (1+s)u_x^+\sigma_x^+
    -
    su_x^-\sigma_x^-,
\end{equation}
or equivalently, for every branch with $p_x>0$,
\begin{equation}
    \zeta_x
    =
    \left(
        1+\frac{su_x^-}{p_x}
    \right)\sigma_x^+
    -
    \frac{su_x^-}{p_x}\sigma_x^-,
\end{equation}
using the fact that $p_x=(1+s)u_x^+-su_x^-$. This is a feasible standard-robustness decomposition of $\zeta_x$ with respect to $\operatorname{STAB}_B$, and hence
\begin{equation}
    p_x R_{\mathrm s}^{\operatorname{STAB}_B}(\zeta_x)
    \leqslant
    su_x^-,
\end{equation}
and summing over all outcomes gives
\begin{equation}
    \sum_xp_x
    R_{\mathrm s}^{\operatorname{STAB}_B}(\zeta_x)
    \leqslant
    s
    \sum_xu_x^-=s=
    R_{\mathrm s}^{\mathcal Q_{B|A}^{(\psi^a)}}(\omega^{AB}).
\end{equation}
Also, we have used the the trace-preserving property of $\sum_x\Lambda_x^\psi$ to conclude that $\sum_xu_x^-=1$. This completes the proof of Theorem~\ref{thm:Steleport}.

\end{document}